\documentclass[12pt]{article}

\usepackage{amsmath}
\usepackage{amsfonts}
\usepackage{amssymb}
\usepackage{graphicx}
\usepackage{grffile}
\usepackage{subfig}
\usepackage{hyperref}
\usepackage{braket}
\usepackage{xcolor}
\usepackage{cite}
\usepackage{dsfont}
\usepackage{float}
\usepackage{ulem}
\usepackage{breqn}
\usepackage{multirow}
\allowdisplaybreaks

\def\be{\begin{equation}}
\def\ee{\end{equation}}
\def\bea{\begin{eqnarray}}
\def\eea{\end{eqnarray}}
\def\nn{\nonumber}

\begin{document}
\thispagestyle{empty}

\begin{center}
{\LARGE\bf{Modular linear differential equations for four-point sphere conformal blocks}}
\bigskip

{\large Ratul Mahanta}\,$^{a}$ \, \, and \, \, {\large Tanmoy Sengupta}\,$^{b,c}$\\

\bigskip 
\bigskip
 
{\small
$^a$ INFN, Sezione di Bologna, viale Berti Pichat 6/2, 40127 Bologna, Italy \\[3mm]
$^b$ The Institute of Mathematical Sciences, IV Cross Road, C.I.T. Campus, Taramani, Chennai 600113, India \\[3mm]
$^c$ Homi Bhabha National Institute, Training School Complex, Anushakti Nagar, Mumbai 400094, India \\[2mm]
}
\end{center}

\begin{center}
\it{mahanta@bo.infn.it, tsengupta@imsc.res.in}
\end{center}
\bigskip 

\begin{center} 
{\bf Abstract} 
\end{center}

\begin{quotation}
\noindent 
{\small
We construct modular linear differential equations (MLDEs) w.r.t. subgroups of the modular group whose solutions are Virasoro conformal blocks appearing in the expansion of a crossing symmetric 4-point correlator on the sphere. This uses a connection between crossing transformations and modular transformations. We focus specifically on second order MLDEs with the cases of all identical and pairwise identical operators in the correlator. The central charge, the dimensions of the above operators and those of the intermediate ones are expressed in terms of parameters that occur in such MLDEs. In doing so, the $q$-expansions of the solutions to the MLDEs are compared with those of Virasoro blocks; hence, Zamolodchikov's elliptic recursion formula provides an important input. Using the actions of respective subgroups, bootstrap equations involving the associated 3-point coefficients have been set up and solved as well in terms of the MLDE parameters. We present explicit examples of MLDEs corresponding to BPZ and novel non-BPZ equations, as well as unitary and non-unitary CFTs.
}
\end{quotation}

\newpage
\tableofcontents
\setcounter{footnote}{0}
\setcounter{page}{1}

\section{Introduction and summary}
\label{sec:intro}

There are infinite-dimensional local algebras in 2 dimensions, having Virasoro algebra as a subalgebra. The conformal blocks on the sphere are determined by these algebras at generic central charge $c$ and operator dimensions $\{h_a,h^{\rm int}_p\}$, where $h_a$ correspond to external operators $O_a$ and $h^{\rm int}_p$ to the intermediate operator $O^{\rm int}_p$. On the other hand, conformal field theories (CFTs) are specified by consistent CFT data namely the central charge, the operator dimensions in the spectrum and the 3-point structure constants. The latter occurs in the conformal block expansion coefficients of the correlators. As one of the consistency criteria we require crossing symmetry of the correlators on the sphere, and to set up such bootstrap equations we need to know the matrices connecting the conformal blocks at two singularities in the cross-ratio $x$. In plenty of cases, blocks have been obtained in closed forms as solutions to differential equations satisfied by the correlators, and then the said connection matrices have been obtained, e.g. \cite{BPZ1984, KZ1984, FLNO2009, Francesco1997}. They have employed null states w.r.t. the local algebra to derive such differential equations; however, see the appendix of \cite{MMS1988c} for a discussion on general construction of such differential equations on the sphere.

This paper discusses the construction of differential equation for the conformal blocks that appear in the expansion of a crossing symmetric 4-point sphere correlator. Crossing symmetry translates as certain modular properties of the correlator under a well-known map \cite{Zamolodchikov1987} between the cross-ratio space and the complex upper half plane $\mathbb{H}_+$ \cite{MMN2016, MM2019, CGL2020}. This allows us to write modular linear differential equations (MLDEs) in $\tau\in\mathbb{H}_+$ w.r.t. some subgroup $\tilde{\Gamma}$ of the modular group for the conformal blocks appearing in the correlator. We determine $\tilde{\Gamma}$ based on whether some operators in the correlator are identical or not. By construction the order of the differential equation is equal to the number of blocks appearing in the expansion of the correlator. Since we do not use null states, our construction is general and includes null state equations as subcases. This construction is in the spirit of \cite{MMS1988a, MMS1988b}, which wrote MLDEs for the characters in rational CFTs in keeping with the modular invariance of the torus partition function formed by the characters.\footnote{See \cite{HM2016a, GHM2016, HM2016b, CM2019a, CM2019b, MNS2021, MPS2020, DGS2021, DGM2022, MR2022} for recent works along this line.} However, the group does not have to be the full modular group in our case. And, we do not need to assume a finite number of operators in the spectrum; we only need to assume a finite number of conformal blocks in the expansion of a correlator $\langle O_1O_2O_3O_4\rangle$ in consideration. The most general scenario for the latter to happen is as follows.\footnote{An integral version can also be written for this scenario resulting in finitely many blocks in the expansion of the correlator $\langle O_1O_2O_3O_4\rangle$.}
\be
  O_1\times O_2=\sum_{p=1}^N O^{\rm int}_{p}+ \sum_q \Xi^{\rm int}_{q},\qquad O_3\times O_4=\sum_{p=1}^N O^{\rm int}_{p}+ \sum_r \tilde{\Xi}^{\rm int}_{r}\,,
\label{eq:genfusionN}
\ee
where in the fusion rules $O_a$ are the 4 external operators and  ``int'' denotes the intermediate operators. And, the operators $\Xi^{\rm int}_{q}$ are all distinct from $\tilde{\Xi}^{\rm int}_{r}$.

Although $\tilde{\Gamma}$ differs depending on whether some or all of the external operators are identical or not, they all contain $\Gamma(2)$ as a subgroup. Thus, we begin with an ansatz for the MLDE w.r.t. $\Gamma(2)$ that introduces a set of parameters $\vec{\alpha}$. Modular forms w.r.t. $\Gamma(2)$ can be expanded in terms of specific powers of the Jacobi theta functions, and $\vec{\alpha}$ are the expansion coefficients. The locus on the parameter space is then found so that it forms an MLDE w.r.t. $\tilde{\Gamma}$. The rationale behind such a construction is as follows. We strip out a specific factor from the correlator that encodes the statistics in 2 dimensions, following \cite{MMN2016}. The crossing symmetry of the stripped correlator translates as the invariance under $\tilde{\Gamma}$-actions on $\tau$. The holomorphic blocks that contribute to the expansion of the stripped correlator are therefore generally allowed to transform as a vector valued modular function under $\tilde{\Gamma}$, and should so satisfy such an MLDE. In particular, the $\Gamma(2)$-actions on $\tau$ generate closed loops circling $x=0,1$. Thus, the $\Gamma(2)$-invariance is interpreted as the single-valuedness of the stripped correlator. Since the statistics are already encoded in the stripped out factor, this should be the case.

In this paper, we further tune $\vec{\alpha}$ such that the solutions to above MLDE are Virasoro conformal blocks. We organise our program as the following series of steps. Solving the MLDE, we obtain the $q$-expansion coefficients of trial solutions at generic $\vec{\alpha}$. The $q$-expansions of the Virasoro blocks can be computed using Zamolodchikov's elliptic recursion relation \cite{Zamolodchikov1987}. We identify $c,h_a,h^{\rm int}_p$ in terms of $\vec{\alpha}$ by comparing the leading behaviour as $q\to0$ and the first few coefficients in the trial solutions with those of Virasoro blocks. The parameter space is then constrained such that the higher coefficients also match. In specific examples, we check it numerically to a sufficiently high order in the $q$-expansions. In latter checks, we numerically produce the $q$-expansion coefficients in the solutions using known $q$-expansions of the modular forms and solving the MLDE recursively. On the other hand, we use a Mathematica code published in \cite{CHKL2017} for the numerical computations of $q$-expansion coefficients in the Virasoro blocks. On the surviving parameter space $\vec{\alpha}$, the solutions to MLDE are 4-point Virasoro blocks. We discuss the (hyper-)surfaces in this parameter space on which the MLDE corresponds to BPZ equations and the possibilities of new equations. Using the $\tilde{\Gamma}$-actions, bootstrap equations involving the associated 3-point coefficients have been set up and solved as well in terms of the MLDE parameters $\vec{\alpha}$. For above implementations, we particularly focus on second order MLDEs with the cases of all identical operators and pairwise identical operators in the correlator. Let us summarize our main results in the following.
\begin{itemize}
  \item{In the case of all identical primaries in the correlator and 2 intermediate ones, one of which is identity and the other with dimension greater than zero, we produce a MLDE w.r.t. the full modular group with a single parameter whose solutions are the 4-point Virasoro blocks. In this case, the MLDE always corresponds to a second order BPZ equation. For scalar operators, the square of the associated non-trivial 3-point coefficient is expressed in terms of the MLDE parameter. Correlator of 4 spinning operators vanishes in this case.}
  
  \item{We also present explicit examples of MLDEs in both unitary and non-unitary theories in the above case. By testing various unitarity bounds, we constrain the parameter space. In the present case, the non-negativity of the $q$-expansion coefficients of the pillow blocks (which are related to the Virasoro blocks) applies as unitarity bound \cite{MSZ2015,Yin2017}. We provide evidence that this is not a sufficient condition for unitarity.}
  
  \item{In the case of pairwise identical primaries in the correlator and 2 intermediate ones, one of which is identity and the other with dimension greater than zero, starting from some specific examples of MLDEs (obtained by a computer-assisted search) we use linear perturbations around them to generate (infinite-)family of MLDEs w.r.t. a subgroup of the modular group whose solutions are the 4-point Virasoro blocks. The product of the associated non-trivial 3-point coefficients is expressed in terms of the MLDE parameters, confining us to scalar operators. We present some MLDEs that do not correspond to BPZ equations. These are new second order differential equations whose solutions are Virasoro blocks. These specific cases are non-unitary.}
  
  \item{We also discuss the importance to study MLDEs with meromorphic forms as coefficients, since by tuning the parameters solutions to such MLDEs may as well give 4-point Virasoro blocks. We present one such example where although the differential equation has additional singularities than the expected ones the Virasoro blocks are free from them.}
\end{itemize}

As it stands now, our program focuses on constructing differential equations for conformal blocks on a basis that is more general than the occurrence of null states, rather than classifying CFTs.\footnote{However, a classification program in the context of MLDEs for blocks may also be initiated by considering correlators of the intermediate operators (which we obtain using our second order MLDEs) with those of the external ones, and writing down MLDEs for the blocks of these new correlators to see the dimensions of possible new(if any) intermediate operators. In turn, we advance toward the full spectrum of the CFT. But, the complexity lies in allowing higher order MLDEs for the blocks in the new correlators.} We do not bear any expectations regarding integrality of the $q$-expansion coefficients in the conformal blocks, unlike characters, which makes our program challenging. In fact, that is another place where our program differs from that of \cite{MMS1988a, MMS1988b}. In the latter demanding integrality of the coefficients greatly helps to obtain a finite number of values of the MLDE parameter, whereas we get connected region(s) in the parameter space $\vec{\alpha}$ where our required MLDEs exist. Nonetheless, we have reported whether the coefficients are non-negative integers or not in each of our examples presented, keeping in mind that the cases with integer coefficients may be suitable for a sphere-torus correspondence as recently reported in \cite{CGL2020}.

This paper is organized as follows. In section \ref{sec:review}, we review the relationship between crossing symmetry and modular transformations. In section \ref{sec:MLDEsph}, we formulate our program of the construction of MLDEs for the Virasoro blocks appearing in the expansion of a crossing symmetric 4-point correlator on the sphere. Section \ref{sec:casestudies} deals with the studies of second order MLDEs with section \ref{sec:allidenO} focusing on the case of all identical external operators and section \ref{sec:pairwiseid} on the pairwise identical external operators. We end by discussing potential interesting generalisations in section \ref{sec:discuss}.

\section{Review}
\label{sec:review}

Under a suitable map between the cross-ratio space and the complex upper half plane, crossing symmetry of a 4-point correlator on the sphere translates as certain modular transformation properties. In this section, we review the necessary elements of this relationship. For more details, see \cite{Zamolodchikov1987, MMN2016, MM2019, CGL2020}.

\subsection{Stripped correlators}
\label{sec:stripcorrel}

For 4 (quasi-)primary operators, \cite{MMN2016} introduced stripped correlators whose transformations under crossing are simple in the sense explained below.

Let $O_1(z_1,\bar{z}_1), \dots ,O_4(z_4,\bar{z}_4)$ be 4 (quasi-)primary operators located on the sphere. A 4-point correlator of these operators is denoted as
$$\langle O_a(z_a,\bar{z}_a) O_b(z_b,\bar{z}_b) O_c(z_c,\bar{z}_c) O_d(z_d,\bar{z}_d)\rangle$$
where $(abcd)$ is a permutation of $(1234)$. It is a certain function of the operator locations and conformal dimensions, which remains invariant up to a phase if we reorder the operators inside the angle brackets $\langle\cdots\rangle$. The latter property is referred to as crossing symmetry and the said phase encodes the statistics of the operators in 2 dimensions. 

Let us label the 4 places from left to right inside the angle brackets $\langle\cdots\rangle$ by $i=1,\dots,4$. With each ordering of the 4 operators inside  the angle brackets, we associate a cross-ratio of their locations, see \eqref{eq:defcrossratio}. Consider any given ordering of the 4 operators, w.r.t. which let $(z(i),\bar{z}(i))$ and $(h(i),\bar{h}(i))$ respectively denote the location and the dimension of the operator that sits at the $i$-th place. The corresponding correlator can be factored into 2 components, one of which is a function of the associated cross-ratio and the other satisfies the Ward identities for Global conformal symmetry. I.e.,
\begin{align}
  &\big\langle \underbrace{}_{\color{red}1} \underbrace{}_{\color{red}2} \underbrace{}_{\color{red}3} \underbrace{}_{\color{red}4} \big\rangle= \text{function of cross-ratio} \times \underbrace{\prod_{i<j=2}^4(z_{ij}^{\mu_{ij}}\cdot\bar{z}_{ij}^{\mu_{ij}})}_{\equiv G_0}\,, \nn \\
  &z_{ij}=z(i)-z(j),\quad \mu_{ij}=\frac{1}{3}\sum_{k=1}^4h(k)-h(i)-h(j),\quad \bar{\mu}_{ij}=\frac{1}{3}\sum_{k=1}^4\bar{h}(k)-\bar{h}(i)-\bar{h}(j)\,, \nn \\
  &\text{cross-ratio}\equiv\frac{\big(z(1)-z(2)\big)\big(z(3)-z(4)\big)}{\big(z(1)-z(3)\big)\big(z(2)-z(4)\big)}\ .
\label{eq:defcrossratio}
\end{align}
A specific solution to the global Ward identities is used above which we denote by $G_0$. Under permutations of operators inside the angle brackets, $G_0$ picks the phase suited to the statistics of the operators. E.g., it remains invariant for scalar operators.

W.r.t. above rule for cross-ratio, the stripped correlators $G_{abcd}$ as functions of cross-ratios are defined as
\begin{align}
  &\big\langle \underbrace{O_a(z_{a},\bar{z}_{a})}_{\color{red}1}\underbrace{O_b(z_{b},\bar{z}_{b})}_{\color{red}2}\underbrace{O_c(z_{c},\bar{z}_{c})}_{\color{red}3}\underbrace{O_d(z_{d},\bar{z}_{d})}_{\color{red}4} \big\rangle=\ G_{abcd}\big(x_{abcd},\bar{x}_{abcd}\big) \times G_0\,, \nn \\
  &x_{abcd} = \frac{(z_a-z_b)(z_c-z_d)}{(z_a-z_c)(z_b-z_d)},\quad (abcd)=\text{permutation of}\ (1234)\,.
\label{eq:strippedcorDef}
\end{align}
Now, denoting $x=x_{1234}$ crossing symmetry relates them as follows\footnote{Here, $x$ lives on a 3-punctured sphere which we call as the cross-ratio space.}
\begin{align}
  {\text{\it non-trival}:}\quad G_{1234}(x,\bar{x})& = G_{1243}(\frac{x}{x-1},\frac{\bar{x}}{\bar{x}-1}) = G_{4231}(\frac{1}{x},\frac{1}{\bar{x}}) = G_{3214}(1-x,1-\bar{x}) \nn \\
  & = G_{3241}(\frac{x-1}{x},\frac{\bar{x}-1}{\bar{x}}) = G_{4213}(\frac{1}{1-x},\frac{1}{1-\bar{x}})\,, \nn \\
  {\text{\it trival}:}\quad G_{1234} (x, \bar{x}) &= G_{2143} (x, \bar{x}) = G_{3412} (x, \bar{x}) = G_{4321} (x, \bar{x})\,.
\label{eq:crossrel}
\end{align}
Furthermore, each $G_{abcd}(x,\bar{x})$ is single-valued in $x$.\footnote{Here, we mean on the slice $\bar{x}=x^*$ with $x^*$ denoting the complex conjugate of $x$.} Only 6 among all the $G_{abcd}$ are independent. From them all the others can be obtained using the trivial relations in \eqref{eq:crossrel}.

We emphasize that the specific selection of $G_0$ results in the aforesaid simple crossing transformations of the stripped correlators. There are no $x$-dependent prefactors in any of the equalities in \eqref{eq:crossrel}, in contrast to \cite{BPZ1984, Francesco1997}. They reveal the role of the full modular group, as discussed in the following subsection. However, the role of modular $S$-transformation is observable with other choices of $G_0$ and has long been studied; for recent considerations, see \cite{DDP2017, DDR2020, CKM2021, Besken2021}.

We also note that
\begin{align}
  \langle O_1(\infty) O_2(1) O_3(x) O_4(0) \rangle&\equiv \lim_{z_1,\bar{z}_1\to\infty}z_1^{2h_1}\bar{z}_1^{2\bar{h}_1}\langle O_1(z_1,\bar{z}_1)O_2(1,1)O_3(x,\bar{x})O_4(0,0)\rangle \nn \\
  &=\ x^{\mu_{34}}(1-x)^{\mu_{23}}  \bar{x}^{\bar{\mu}_{34}}(1-\bar{x})^{\bar{\mu}_{23}} G_{1234}(x,\bar{x})\,.
\label{eq:asympcor}
\end{align}
The holomorphic conformal blocks that contribute to the stripped correlator $G_{1234}(x,\bar{x})$ exhibit the following leading behaviour as $x\to0$.\footnote{Since the holomorphic conformal blocks that contribute to $\langle O_1(\infty) O_2(1) O_3(x) O_4(0) \rangle$ go as $x^{h^{\rm int}_{p}-h_3-h_4}$ in the limit $x\to0$ \cite{BPZ1984, Francesco1997}.}
\be
  F_{p}(x)=x^{h^{\rm int}_{p}-\frac{\mathfrak{H}}{3}}\left(1+\mathcal{O}(x)\right),\quad \mathfrak{H}=h_1+\cdots+h_4\,,
\ee
where $h^{\rm int}_{p}$ denotes the dimension of the intermediate operator $O^{\rm int}_{p}$ that appears in the fusion channel: $O_3\times O_4=\sum_p O^{\rm int}_{p}$. As opposed to the intermediate ones, we call the operators which sit inside the angle brackets as external operators.

\subsection{Modular properties of stripped correlators}
\label{sec:modstripcor}

Consider the map from the complex upper half plane $\mathbb{H}_+$ to the cross-ratio space\footnote{We take the cross-ratio space as $\mathbb{C}\setminus\{0,1\}$ setting the 3 punctures at $0,1,\infty$ using global conformal transformations.} given by the elliptic lambda function as follows.\footnote{$\theta_2(\tau),\theta_3(\tau),\theta_4(\tau)$ are the Jacobi theta functions related by Jacobi Identity: $\theta_4(\tau)^4=\theta_3(\tau)^4-\theta_2(\tau)^4$. See \cite{Handbooks} for further details about $\lambda(\tau),\theta_r(\tau)$.}
\begin{align}
  &x=\lambda(\tau)= \left(\frac{\theta_2(\tau)}{\theta_3(\tau)}\right)^4,\quad \tau \in \mathbb{H}_+\,, \nn \\
  &\theta_2(\tau)=\sum_{n\in \mathbb{Z}}e^{\pi i \tau (n+\frac{1}{2})^2},\quad\theta_3(\tau)=\sum_{n\in \mathbb{Z}}e^{\pi i \tau n^2}\,.
\label{eq:mapUHPx}
\end{align}
The action of the modular group $PSL(2,\mathbb{Z})$ on $\tau$ generates all the crossing transformations on $x$. In particular, under the actions of the 2 generators of the modular group which are $T:\tau \to \tau + 1,\ S:\tau \to -\frac{1}{\tau}$, the images on the cross-ratio space respectively have the transformations: $T\cdot x=\frac{x}{x-1},\ S\cdot x=1-x$.

$\lambda(\tau)$ remains invariant under the index 6 normal subgroup $\Gamma(2)$ of the modular group generated by $T^{2}$ and $ST^{2}S$. In particular, $T^{2},ST^{2}S$ generate simple loops around $x=0,1$ respectively. Consequently, $\Gamma(2)$ plays a role in single-valuedness of each $G_{abcd}(x,\bar{x})$ in $x$ as discussed below.

Arrange any 6 independent $G_{abcd}$ in a column and consider it to be a vector-valued modular function (vvmf)\footnote{\label{fn:vvmf}More generally, w.r.t. a subgroup $\Gamma\subset PSL(2,\mathbb{Z})$ a vector-valued modular form $\vec{V}$ of weight $k$ is defined as \cite{Gannon2013}
$$\vec{V}(\frac{a\tau+b}{c\tau+d})=\sigma(\begin{pmatrix}a & b\\c & d\end{pmatrix})(c\tau+d)^k\vec{V}(\tau),\quad \begin{pmatrix}a & b\\c & d\end{pmatrix}\in\Gamma\,,$$
with $\sigma$ as linear a representation of $\Gamma$. Hereafter, the shorthand vvmf refers to a vector-valued modular form of weight $0$ w.r.t. some subgroup $\Gamma$. The subgroup should be clear from the context unless specified.} w.r.t. $PSL(2,\mathbb{Z})$ on $\mathbb{H}_+$ as follows.
\begin{align}
  &\vec{G}(\tau,\bar{\tau})=\begin{pmatrix} G_{1234}(\tau, \bar{\tau})\\G_{2134}(\tau, \bar{\tau})\\G_{4132}(\tau, \bar{\tau})\\G_{1432}(\tau, \bar{\tau})\\G_{2431}(\tau, \bar{\tau})\\G_{4231}(\tau, \bar{\tau}) \end{pmatrix}:\quad \vec{G}(\gamma\tau,\gamma\bar{\tau})= \sigma(\gamma) \cdot \vec{G}(\tau,\bar{\tau}) \text{, } \quad \gamma \in PSL(2,\mathbb{Z})\,, \nn \\
  &\sigma(S)=
  \begin{pmatrix}
    0 & 0 & 0 & 1 & 0 & 0\\
    0 & 0 & 0 & 0 & 1 & 0\\
    0 & 0 & 0 & 0 & 0 & 1\\
    1 & 0 & 0 & 0 & 0 & 0\\
    0 & 1 & 0 & 0 & 0 & 0\\
    0 & 0 & 1 & 0 & 0 & 0
  \end{pmatrix},\quad
  \sigma(T)=
  \begin{pmatrix}
    0 & 1 & 0 & 0 & 0 & 0\\
    1 & 0 & 0 & 0 & 0 & 0\\
    0 & 0 & 0 & 1 & 0 & 0\\
    0 & 0 & 1 & 0 & 0 & 0\\
    0 & 0 & 0 & 0 & 0 & 1\\
    0 & 0 & 0 & 0 & 1 & 0
  \end{pmatrix}.
\label{eq:stripcorgenmod}
\end{align}
More precisely, matrices $\sigma(\gamma)$ form a linear representation of $PSL(2,\mathbb{Z})/\Gamma(2)$. The chosen matrices $\sigma(S), \sigma(T)$ ensure the non-trivial crossing transformations in \eqref{eq:crossrel}. Also, we have $\sigma(T^2)=\sigma(ST^2S)=\mathds{1}$. As a result, each component of $\vec{G}(\tau, \bar{\tau})$ has $\Gamma(2)$-invariance, which ensures its single-valuedness in $x=\lambda(\tau)$.

A few comments are in order. In case all the 4 external operators are distinct, all the 6 components of $\vec{G}(\tau,\bar{\tau})$ are independent and each component is invariant under only $\Gamma(2)$. But, $\sigma(S),\ \sigma(T)$ are reducible when some or all of the 4 external operators are identical. In the latter case, each component $G_a(\tau,\bar{\tau})$ is invariant under a bigger subgroup $\tilde{\Gamma}_a$ that contains $\Gamma(2)$. E.g., when there are 4 identical operators, $\vec{G} (\tau,\bar{\tau})$ has only 1 independent component which is modular invariant. When there are 2 pairs of identical operators, $\vec{G}(\tau,\bar{\tau})$ has 3 independent components; the stabilizer subgroup of each of them contains $\Gamma(2)$. In particular, the following representation will be useful for the present work.
\begin{align}
  &O_1=O_2\equiv O_L,\ O_3=O_4\equiv O_R\,, \nn \\
  &\vec{G}(\tau,\bar{\tau})=\begin{pmatrix} G_{LLRR}(\tau, \bar{\tau})\\G_{LRLR}(\tau, \bar{\tau})\\G_{LRRL}(\tau, \bar{\tau}) \end{pmatrix},\quad
  \sigma(S)=
  \begin{pmatrix}
    0 & 0 & 1\\
    0 & 1 & 0\\
    1 & 0 & 0
  \end{pmatrix},\quad
  \sigma(T)=
  \begin{pmatrix}
    1 & 0 & 0\\
    0 & 0 & 1\\
    0 & 1 & 0
  \end{pmatrix}\,.
\end{align}
In this case, the stabilizer of the vector $(1\ 0\ 0)^t$ is the subgroup of $PSL(2,\mathbb{Z})$ generated by $T,ST^2S$.

Here, we tabulate the possible cases along with respective stabilizer subgroups as per our cross-ratio convention and \eqref{eq:stripcorgenmod}.
\begin{table}[H]
\begin{center}
\resizebox{\columnwidth}{!}{%
        \begin{tabular}{|c|c|c|}
                \hline
                 & Independent stripped correlators & Generators of stabilizer subgroup \\
                \hline
                {All identical operators} & {$G_{OOOO}$} & {$S,\ T$}  \\
                \hline
                \multirow{3}{*}{Pairwise identical operators} & {$G_{LLRR}$} & {$T,\ ST^{2}S$} \\
                & {$G_{RLRL}$} & {$S,\ T^{2}$}\\
                & {$G_{LRRL}$} & {$T^2,\ STS$}\\
                \hline
                \multirow{3}{*}{3 identical operators} & {$G_{LRRR}$} & {$S,\ T^{2}$} \\
                & {$G_{RLRR}$} & {$T^2,\ STS$}\\
                & {$G_{RRRL}$} & {$T,\ ST^{2}S$}\\
                \hline
                {All distinct operators} & {$\begin{aligned} &\ \ \ \ \ G_{abcd}\ ,\\[-7pt] (abcd)= &\text{permutations of}\ (1234) \end{aligned}$} & {$T^2,\ ST^2S$}  \\
                \hline
        \end{tabular}
}
\end{center}
\caption{List of possible stabilzer subgroups}
\label{tab:stabilizer}
\end{table}

\section{MLDE for sphere blocks and bootstrap}
\label{sec:MLDEsph}

Here, we consider the sphere correlator: $\langle O_1(z_1,\bar{z}_1)O_2(z_2,\bar{z}_2)O_3(z_3,\bar{z}_3)O_4(z_4,\bar{z}_4)\rangle$. From \eqref{eq:strippedcorDef} and \eqref{eq:mapUHPx}, the associated stripped correlator is $G_{1234}(\tau,\bar{\tau})$, which is the first component of the vvmf $\vec{G}(\tau,\bar{\tau})$ in \eqref{eq:stripcorgenmod}. It has invariance under a subgroup $\tilde{\Gamma}_1\supseteq\Gamma(2)$ of the modular group, as discussed in section \ref{sec:modstripcor}. We confine ourselves to the case where a finite number of intermediate operators occur in the s-channel OPE $O_3\times O_4$.\footnote{More generally, the assumption \eqref{eq:genfusionN} suffices for our analysis.} We denote the number by $N$.

Let the $N$ intermediate operators that contribute to the s-channel conformal block expansion of the stripped correlator $G_{1234}(\tau,\bar{\tau})$ be: $O_{p(k)},\ k=1,\dots,N$ with respective dimensions $(h_{p(k)},\bar{h}_{p(k)})$. Since the real $(c,h)$ plane is the primary focus of this article, we take $h_{p(k)}$ to be ordered as
\be
  h_{p(1)} \leq h_{p(2)} \leq \cdots \leq h_{p(N)}\,,
\label{eq:hporder}
\ee
without loss of generality. Now, we have
\be
  G_{1234}(\tau,\bar{\tau})=\sum_{k=1}^N C_{12p(k)}C_{34p(k)}F_{p(k)}(\tau,\bar{\tau})\,,
\ee
where $C_{ijp(k)}$ are the 3-point coefficients and $F_{p(k)}(\tau,\bar{\tau})$ are the non-holomorphic conformal blocks respective to operators $O_{p(k)}$. Each $F_{p(k)}(\tau,\bar{\tau})$ further factorizes as
\be
  F_{p(k)}(\tau,\bar{\tau})=F_{p(k)}(\tau){\bar{F}}_{p(k)}(\bar{\tau})\,,
\label{eq:HoloAntiFac}
\ee
where ${\bar{F}}_{p(k)}(\bar{\tau})$ is obtained with replacing $\{h_1,\dots,h_4,h_{p(k)},\tau\}$ by $\{\bar{h}_1,\dots,\bar{h}_4,\bar{h}_{p(k)},\bar{\tau}\}$ in the holomorphic block $F_{p(k)}(\tau)$.

The holomorphic blocks $F_{p(k)}(\tau)$ mix among themselves under the action of $\tilde{\Gamma}_1$ on the complex upper half plane $\mathbb{H}_+$, while $G_{1234}(\tau,\bar{\tau})$ is an invariant.\footnote{In particular, mixing under $\Gamma(2)$ gives monodromies of the holomorphic blocks around $x=0,1$. In contrast to the non-trivial monodromies around $x=1$, those around $x=0$ are only phases.} I.e., the blocks $F_{p(k)}(\tau)$ can be arranged in a row which transform as a vvmf w.r.t. $\tilde{\Gamma}_1$ as follows.\footnote{For definition, see footnote \ref{fn:vvmf}.}
\begin{align}
  &F_{p(k)}(\gamma\tau)=\sum_{k'}F_{p(k')}(\tau)M(\gamma)_{p(k')p(k)},\quad \gamma\in\tilde{\Gamma}_1\supseteq\Gamma(2)\,, \nn \\
  &M(\gamma_2\cdot\gamma_1)=M(\gamma_1)\cdot M(\gamma_2)\,,
\label{eq:blocksvvmf}
\end{align}
where matrices $M(\gamma)$ form a linear representation of $\tilde{\Gamma}_1$. Now, we regard the $N$ holomorphic blocks $F_{p(k)}(\tau)$ to be the independent solutions of a linear differential equation in $\tau$ of order $N$. And, \eqref{eq:blocksvvmf} requires the differential equation to be invariant under $\tilde{\Gamma}_1$ i.e. an MLDE w.r.t.  $\tilde{\Gamma}_1$. Since $\tilde{\Gamma}_1\supseteq\Gamma(2)$\footnote{Where equality holds only when all the external operators are distinct.}, an MLDE w.r.t. $\tilde{\Gamma}_1$ is always an MLDE w.r.t. $\Gamma(2)$, but the converse is not true. Below, we take the strategy to start with the general form of an MLDE w.r.t. $\Gamma(2)$ which introduces a set of parameters. Then, we find the locus on the parameter space such that the differential equation becomes an MLDE w.r.t. $\tilde{\Gamma}_1$.

To begin, the most generic ansatz of aforesaid differential equation that can be written for the blocks is given by\footnote{In general, the differential operator in l.h.s. of \eqref{eq:genmlde} may act on a vector-valued modular form of weight $k$ w.r.t. $\Gamma(2)$. But, for all our purposes, it operates on a vvmf.}
\be
  D^{N}\vec{F}+\sum_{s=1}^{N-1}\phi_{s}(\tau)D^{s}\vec{F}+\phi_{0}(\tau)\vec{F}=0,\quad \vec{F}=\left(F_{p(1)}(\tau)\ \cdots\ F_{p(N)}(\tau)\right),
\label{eq:genmlde}
\ee
where $\phi_{r}(\tau)$ are some (meromorphic-)modular forms of weight $2(N-r)$ w.r.t. $\Gamma(2)$ and $D^r$ stands for the $r$ successive operations of the Serre derivative $D$. At each operation, $D$ depends on the weight $k$ of the form on which it operates as follows.
\be
  D=\frac{d}{d\tau}-\frac{i\pi k}{6}E_2(\tau)\,,
\ee
where $E_2(\tau)$ is the second Eisenstein series.\footnote{Note the modular transformation property of the second Eisenstein series \cite{Handbooks}
$$E_2(\frac{a\tau+b}{c\tau+d})=(c\tau+d)^2E_2(\tau)+\frac{6}{i\pi}c(c\tau+d),\quad \begin{pmatrix}a & b\\c & d\end{pmatrix}\in PSL(2,\mathbb{Z})\,.$$} E.g., $D\vec{F}=\frac{d}{d\tau}\vec{F}$ as $k=0$ in this case, while $D^2\vec{F}=\left(\frac{d}{d\tau}-\frac{i\pi}{3}E_2(\tau)\right)\frac{d}{d\tau}\vec{F}$, etc.

A few comments are in order. Firstly, the objective of using the Serre derivate is to produce a $k+2$ form operating on a $k$ form w.r.t. $\Gamma(2)$. Unlike the case of $PSL(2,\mathbb{Z})$, as there exist forms of weight 2 w.r.t. $\Gamma(2)$ they may be added to modify the Serre derivative as follows.
\be
  \tilde{D}=D+\tilde{\alpha}_1\theta_2(\tau)^4+\tilde{\alpha}_2\theta_3(\tau)^4,\quad \tilde{\alpha}_1,\tilde{\alpha}_2\in\mathbb{C}\,.
\ee
However, an MLDE written using $\tilde{D}$ can always be brought to the form \eqref{eq:genmlde}. Thus, we continue to use simply $D$. Secondly, the requirement of \eqref{eq:genmlde} to be an MLDE w.r.t. $\tilde{\Gamma}_1$ ($\neq\Gamma(2)$ when some/all of the operators are identical) restricts the locus of $\phi_{r}(\tau)$ on the space of $2(N-r)$-forms w.r.t. $\Gamma(2)$.

In the following, we organise the present program in a series of steps.\footnote{\label{fn:MLDEantiholblocks}We detail the construction of MLDE for the holomorphic blocks, but similar steps can be followed with dimensions $\{\bar{h}_1,\dots,\bar{h}_4,\bar{h}_{p(k)}\}$ producing the anti-holomorphic blocks at $\tau$ i.e. ${\bar{F}}_{p(k)}(\tau)$. At the end, we evaluate them at $\bar{\tau}(=\tau^*)$. Hereafter, blocks refer to holomorphic blocks; when we need to glue them with anti-holomorphic ones we use specific words.} We apply them to definite cases in the subsequent section.

\vspace{0.3cm}
\noindent \underline{Step 1}: We fix $N\geq2$ in \eqref{eq:genmlde}. We expand each $\phi_{r}(\tau)$ on a given basis of the space of (meromorphic-)modular forms of weight $2(N-r)$ w.r.t. $\Gamma(2)$. The expansion coefficients, collectively called as $\vec{\alpha}$, are complex-valued parameters. However, as the real $(c,h)$ plane is the primary focus of this article, scanning over real $\vec{\alpha}$ will often suffice.

\vspace{0.3cm}
\noindent \underline{Step 2}: Based on which external operators are identical, we determine the subgroup $\tilde{\Gamma}_1$. We find the locus on the parameter space $\vec{\alpha}$ such that each $\phi_{r}(\tau)$ become a form of weight $2(N-r)$ w.r.t. $\tilde{\Gamma}_1$. For the subsequent steps, we stay on this locus.

\vspace{0.3cm}
\noindent \underline{Step 3}: We take trial solutions to \eqref{eq:genmlde} to be of the form
\be
  q^s\sum_{n=0}^\infty a_n q^n=a_0q^s\big(1+\sum_{n=1}^\infty \frac{a_n}{a_0} q^n\big),\quad a_0\neq0,\quad q=e^{i\pi\tau}\,.
\label{eq:trailsolgen}
\ee
The indicial equation is a polynomial equation of degree $N$ in $s$, which we solve to get $N$ exponents in terms of the parameters $\vec{\alpha}$.\footnote{This can always be done analytically for $N<5$.} We narrow the parameter space further such that each of these exponents leads to a series solution consistent with $a_0\neq0$.\footnote{This entails removing points $\vec{\alpha}$ where a pair of exponents are separated by an even integer.} On the surviving parameter space, for each of these exponents $s(\vec{\alpha})$, \eqref{eq:genmlde} being a linear differential equation does not fix $a_0$. However, it gives recursion relation(s) for the higher coefficients $a_n$ in terms of the lower coefficients. The values $s(\vec{\alpha})$ encode information about the dimensions of the intermediate operators, while the coefficients $a_n,\ n\geq2$ encode information about the central charge as well.

The preceding 3 steps work for any 4 external (quasi-)primaries $O_a$, since \eqref{eq:stripcorgenmod} functions. In particular for quasi-primaries, \eqref{eq:genmlde} is supposed to be an MLDE for the global conformal blocks which have no dependencies on central charge. For the subsequent steps, we confine ourselves to Virasoro primaries $O_a$. In addition, we consider the expansion of the stripped correlator in terms of Virasoro blocks.

\vspace{0.3cm}
\noindent \underline{Step 4}: We consider the solutions of \eqref{eq:genmlde} to be 4-point Virasoro blocks. For generic central charge $c$, external operator dimensions $h_a$ and internal dimensions $h_{p(k)}$, the $q$-expansions of respective Virasoro blocks can be computed \cite{Zamolodchikov1987}. We compare them with \eqref{eq:trailsolgen} to relate the quantities $\{c,h_a,h_{p(k)}\}$ with parameters $\vec{\alpha}$ as follows. In terms of Zamolodchikov's $H$-function, the Virasoro blocks that contribute to the stripped correlator are given by\footnote{Note that the $\lambda$-dependent factors multiplying $H$ are in accordance with \eqref{eq:asympcor}, and hence differ from \cite{Zamolodchikov1987}. Also, recall that $\mathfrak{H}=h_1+\cdots+h_4$.}
\begin{align}
  &F_{p(k)}(\tau)=(16 q)^{h_{p(k)}-\frac{c-1}{24}} \lambda(\tau)^{\frac{c-1}{24}-\frac{\mathfrak{H}}{3}} (1-\lambda(\tau))^{\frac{c-1}{24}-\frac{\mathfrak{H}}{3}} \theta_3(\tau)^{\frac{c-1}{2}-4\mathfrak{H}}H(c,h_a,h_{p(k)},q)\,, \nn \\
  &H(c,h_a,h_{p(k)},q)=1+\sum_{n=1}^{\infty}b_n(c,h_a,h_{p(k)})q^n\,,
\label{eq:VBlockH}
\end{align}
where the coefficients $b_n$ can be obtained using Zamolodchikov's elliptic recursion relation, see appendix \ref{app:Hfunction}. $b_1$ is independent of $c$. Employing the respective $q$-expansions of $\lambda(\tau),\theta_3(\tau)$ as given in appendix \ref{app:qexp}, we get
\begin{align}
  F_{p(k)}(\tau)&=(16q)^{h_{p(k)}-\frac{\mathfrak{H}}{3}}\left(1+ b_1 q+ \big(b_2+\frac{1-c+8 \mathfrak{H}}{2}\big)q^2+\big(\frac{1-c+8 \mathfrak{H}}{2}b_1+b_3\big)q^3 +\cdots\right) \nn \\
  &\equiv (16q)^{h_{p(k)}-\frac{\mathfrak{H}}{3}} \bigg(1+\sum_{n=1}^{\infty}f_n(c,h_a,h_{p(k)}) q^n\bigg).
\label{eq:qexpVblockgen}
\end{align}
The $q$-expansion coefficients of $F_{p(k)}$ are linearly related to those of $H$. In fact, the following general patterns arise in \eqref{eq:qexpVblockgen} with some coefficients $\beta_{nr}(c,\mathfrak{H}),\gamma_{nr}(c,\mathfrak{H})$.
\be
  f_{2n-1}=\sum_{r=1}^n\beta_{nr} b_{2r-1},\quad f_{2n}=\gamma_{n0}+\sum_{r=1}^n\gamma_{nr} b_{2r},\quad n\geq1\,,
\label{eq:linrelfnbn}
\ee
which are obtained using the $q$-expansion of $q^{-1}\eta(\tau)$ with known even coefficients $\sigma_{2n}$ while odd ones vanish as follows.\footnote{Below, we use \eqref{eq:VBlockH}, $\lambda=\theta_2^4/\theta_3^4,\ \theta_4^4=\theta_3^4-\theta_2^4$ and $\theta_2\theta_3\theta_4=2\eta^3$, where $\eta(\tau)=q^{\frac{1}{12}}\prod_{r=1}^{\infty}(1-q^{2r})$ is the Dedekind eta function. $\sigma_{2n}$ can be obtained by using \eqref{eq:qexpThetasEta} up to any desired $n$. Thereby, the coefficients $\beta_{nr}(c,\mathfrak{H})$, $\gamma_{nr}(c,\mathfrak{H})$ are known as well for any desired $n,r$.}
\begin{align}
  &(16q)^{-h_{p(k)}+\frac{\mathfrak{H}}{3}}F_{p(k)}=(16q)^{-\delta}\theta_2^{4\delta}\theta_3^{4\delta}\theta_4^{4\delta} H,\quad \delta=\frac{c-1}{24}-\frac{\mathfrak{H}}{3}\,, \nn \\
  &(16q)^{-1}\theta_2^{4}\theta_3^{4}\theta_4^{4}=q^{-1}\eta^{12}=1+\sum_{n=1}^{\infty}\sigma_{2n}q^{2n}\,.
\label{eq:VBHeta}
\end{align}
Note that if the odd coefficients vanish in $H$, the odd coefficients vanish in $F_{p(k)}$. We need to equate quantities in \eqref{eq:qexpVblockgen} to respective quantities in \eqref{eq:trailsolgen} which are obtained by solving \eqref{eq:genmlde}:
\be
\begin{aligned}
  {\text{\it for arbitrary $q$ near $0$}:}\ \ \ &(16q)^{h_{p(k)}-\frac{\mathfrak{H}}{3}}=a_0q^{s(\vec{\alpha})}\,, \\
  {\text{\it for $n\geq1$}:}\ \ \ &f_n(c,h_a,h_{p(k)})=\frac{a_n(\vec{\alpha})}{a_0}\,.
\label{eq:solVsVB}
\end{aligned}
\ee
At this stage, the number of independent CFT quantities $\{c,h_a,h_{p(k)}\}$ is $N+5$ or lower when some of the operators have same dimensions. In order to generate this many independent equations, we use \eqref{eq:solVsVB} with the first few $n$. These in principle can be inverted to express the aforesaid CFT quantities in terms of the parameters $\vec{\alpha}$. Then, plugging these $c(\vec{\alpha}),h_a(\vec{\alpha}),h_{p(k)}(\vec{\alpha})$ we check the required validity of \eqref{eq:solVsVB} for higher $n$, which may further restrict the parameter space $\vec{\alpha}$. In cases where the said inversion and/or the checks for higher $n$ are cumbersome to perform analytically, we resort to numerical techniques.

On the surviving parameter space, the solutions to MLDE \eqref{eq:genmlde} are 4-point Virasoro blocks. We discuss the (hyper-)surfaces in this parameter space on which the MLDE corresponds to BPZ equations and the possibilities of new equations.

\vspace{0.3cm}
\noindent \underline{Step 5}: In terms of the solutions to \eqref{eq:genmlde} (which are identified as Virasoro blocks), the stripped correlator takes the form:
\be
  G_{1234}(\tau,\bar{\tau})=\left(F_{p(1)}(\tau)\ \cdots\ F_{p(N)}(\tau)\right)\cdot \mathcal{C}\cdot \left(\bar{F}_{p(1)}(\bar{\tau})\ \cdots\ \bar{F}_{p(N)}(\bar{\tau})\right)^t,
\ee
where $\mathcal{C}$ is an $N\times N$ matrix containing the associated 3-point coefficients. Now, the $\tilde{\Gamma}_1$-invariance of $G_{1234}$ sets up the bootstrap equations:
\be
  M(\gamma)\cdot \mathcal{C}\cdot \bar{M}(\gamma)^\dagger=\mathcal{C},\quad \gamma\in\tilde{\Gamma}_1\,,
\label{eq:genbootstrap}
\ee
where $\bar{M}$ is obtained from $M$ using the replacement rule stated below \eqref{eq:HoloAntiFac}. Confining $\gamma$ over the generators of $\tilde{\Gamma}_1$ suffices due to the multiplication rule in \eqref{eq:blocksvvmf}. Note that knowing the $q$-expansion of the blocks around $q=0$ is not enough to derive the transformation matrices $M(\gamma)$. Writing down the MLDE \eqref{eq:genmlde} for the blocks is useful in many instances where we can solve it to get the blocks in closed forms. In latter cases, we obtain $M(\gamma)$ in terms of $\vec{\alpha}$ for the holomorphic blocks. For spinning operators, the anti-holomorphic conformal dimension differs from the holomorphic one. Thus, $\{\bar{h}_1,\dots,\bar{h}_4,\bar{h}_{p(k)}\}$ correspond to different values of the parameters denoted by $\vec{\bar{\alpha}}$ such that $c(\vec{\alpha})=c(\vec{\bar{\alpha}})$, and the solutions to \eqref{eq:genmlde} with parameters $\vec{\bar{\alpha}}$ are the anti-holomorphic blocks at $\tau$.\footnote{See footnote \ref{fn:MLDEantiholblocks}.} The complex conjugation acts in \eqref{eq:genbootstrap} as we evaluate the anti-holomorphic blocks at $\bar{\tau}=\tau^*$. We solve the above matrix equations to obtain $\mathcal{C}$ in terms of $\vec{\alpha},\vec{\bar{\alpha}}$. Note that when we deal with only scalar operators, we have $\vec{\alpha}=\vec{\bar{\alpha}}$ and $M=\bar{M}$.

\vspace{0.3cm}
\noindent \underline{Step 6}: We attempt to differentiate unitary CFTs from non-unitary ones on the surviving parameter space $\vec{\alpha}$ and present examples of each. For unitary theories, we should necessarily set $c,h_a,h_{p(k)}\geq0$. In our context, unitarity bounds on the $q$-expansion coefficients of the blocks will be of importance. The only known such bounds are for the case $h_1=h_4,h_2=h_3$, as stated in the following in terms of the quantities used in the present work. Consider the following $q$-expansion coefficients $\tilde{f_n}$ in linear relation to $f_r$.
\begin{align}
  {\text{\it pillow blocks}:}\ \ \ &\tilde{F}_{p(k)}(\tau)\equiv16^{\frac{\mathfrak{H}}{3}-\frac{c}{24}}\eta(\tau)^{4 \mathfrak{H}-\frac{c}{2}}F_{p(k)}(\tau)=(16q)^{h_{p(k)}-\frac{c}{24}}\bigg(1+\sum_{n=1}^{\infty}\tilde{f}_n q^n\bigg)\,, \nn \\
  {\text{\it first few coefficients}:}\ \ \ &\tilde{f}_1=f_1,\quad \tilde{f}_2=f_2+\frac{1}{2} (c-8 \mathfrak{H}),\quad \tilde{f}_3=f_3+\frac{1}{2} (c-8 \mathfrak{H}) f_1\,, \nn \\
  &\tilde{f}_4=f_4+\frac{1}{2} (c-8 \mathfrak{H}) f_2+\frac{1}{8} (c-8 \mathfrak{H}) (c-8 \mathfrak{H}+6),\quad \dots\,, \nn \\
  {\text{\it general patterns}:}\ \ \ &\tilde{f}_{2n-1}=\sum_{r=1}^n\tilde{\beta}_{nr} f_{2r-1},\quad \tilde{f}_{2n}=\tilde{\gamma}_{n0}+\sum_{r=1}^n\tilde{\gamma}_{nr} f_{2r},\quad n\geq1\,.
\label{eq:genpillowblock}
\end{align}
In the above, $\tilde{F}_{p(k)}$ are known as the pillow blocks. The coefficients $\tilde{f}_n$ can be obtained to any desired order $q^n$, using \eqref{eq:qexpVblockgen} and \eqref{eq:qexpThetasEta}, thereby the coefficients $\tilde{\beta}_{nr}(c,\mathfrak{H})$, $\tilde{\gamma}_{nr}(c,\mathfrak{H})$ are known as well. For the case $h_1=h_4,h_2=h_3$, all the $\tilde{f}_n\geq0$ \cite{MSZ2015,Yin2017}, as they can be interpreted as the norms on some Hilbert space.\footnote{The work \cite{MSZ2015} provides an alternative viewpoint on $q$-expansions of Virasoro blocks, under the map from the Riemann sphere with 4 marked points $\infty,1,x,0$ to the ``pillow'' geometry $T^2/\mathbb{Z}_2$ with 4 corners. This enables to uncover the above non-negativity conditions.} In our fourth step, we have obtained $f_r,c,\mathfrak{H}$ in terms of $\vec{\alpha}$. Therefore, $\tilde{f}_r$ can be expressed in terms of $\vec{\alpha}$ using \eqref{eq:genpillowblock}. Hence, the non-negativity of $\tilde{f}_n(\vec{\alpha})$ restricts the parameter space $\vec{\alpha}$ when we have $h_1=h_4,h_2=h_3$ for unitary theories.

In case the Virasoro representations corresponding to $\bar{h}_a,\bar{h}_{p(k)}$ are unitary, similar constraints apply on $\vec{\bar{\alpha}}$. In addition, the reality of the 3-point coefficients associated with the stripped correlator (obtained in terms of $\vec{\alpha},\vec{\bar{\alpha}}$ in our fifth step) may narrow both parameter spaces $\vec{\alpha},\vec{\bar{\alpha}}$ further, in unitary CFTs.

\section{Case studies: second order MLDEs}
\label{sec:casestudies}

In the simplest scenario, in \eqref{eq:genmlde} $\phi_{r}(\tau),\ r=0,\dots,N-1$ are holomorphic forms of respective weights $2(N-r)$ w.r.t. $\Gamma(2)$ on $\mathbb{H}_+$. Since $\theta_2(\tau)^{4a}\theta_3(\tau)^{4b}$ with non-negative integers $a,b$ form the basis of the space of holomorphic $2(a+b)$-forms w.r.t. $\Gamma(2)$ on $\mathbb{H}_+$, we can write
\be
  \phi_r=\sum_{\substack{a,b=0,1,2,\dots\\ a+b=N-r}}\alpha_{ab}\theta_2^{4a}\theta_3^{4b},\quad r=0,\dots,N-1\,,
\label{eq:Gamma2holforms}
\ee
for some coefficients $\alpha_{ab}$.

The present program has been organised as a series of steps in section \ref{sec:MLDEsph}. In this section, we implement this for the case: $N=2$ with holomorphic forms $\phi_0,\phi_1$ of respective weights $4,2$ w.r.t. $\Gamma(2)$. In this case, we write down the following ansatz introducing parameters $\vec{\alpha}=(\alpha_1,\dots,\alpha_5)$, completing \uwave{our first step}.\footnote{The expansion coefficients $\vec{\alpha}$ on the basis \eqref{eq:Gamma2holforms} are chosen in a specific way for later convenience.}
\begin{align}
  &\frac{d^2\vec{F}}{d\tau^2}-\frac{i\pi}{3}E_2(\tau)\frac{d\vec{F}}{d\tau}+i\pi\left((\frac{1}{3}-2\alpha_1)\theta_2(\tau)^4+(\frac{1}{3}-2\alpha_2)\theta_3(\tau)^4\right)\frac{d\vec{F}}{d\tau} \nn \\
  &\qquad\qquad+\pi^2\left(\alpha_3\theta_2(\tau)^8+\alpha_4\theta_2(\tau)^4\theta_3(\tau)^4+\alpha_5\theta_3(\tau)^8\right)\vec{F}=0\,, \nn \\
  &\vec{F}=\left(F_{p(1)}(\tau)\ F_{p(2)}(\tau)\right).
\label{eq:2ndmlde}
\end{align}
As shown in appendix \ref{app:tautox}, \eqref{eq:2ndmlde} leads to the most general second order linear differential equation with 3 regular singular points in variable $x=\lambda(\tau)$.\footnote{Hence, clearly meromorphic $\phi_0,\phi_1$ should result in more and/or new type singularities. See section \ref{sec:discuss}, for a discussion.} In the following two subsections, we implement the remaining steps separately to the cases $O_1=O_2=O_3=O_4$ and $O_1=O_2,\ O_3=O_4$.

\subsection{All identical operators}
\label{sec:allidenO}

Let us take $O_1=O_2=O_3=O_4\equiv O$. Now, we implement \uwave{the second step}. In the present case, $\tilde{\Gamma}_1$ is the full modular group. In order for \eqref{eq:2ndmlde} to be an MLDE w.r.t. full modular group, we must have
\be
  \alpha_1=\alpha_2=\frac{1}{6},\quad \alpha_3=\alpha_5=-\alpha_4\,.
\label{eq:alpharelO}
\ee
The rationale behind this is as follows. The space of holomorphic $(4a+6b)$-forms w.r.t. the modular group is spanned by $E_4(\tau)^aE_6(\tau)^b$ with non-negative integers $a,b$. The Eisenstein series $E_4,E_6$ being forms w.r.t. the modular group are also forms w.r.t. any subgroup without altering respective weights. Thus, they can be expressed in terms of $\theta_2^4,\theta_3^4$. Clearly, there is no holomorphic modular form of weight $2$ and only 1 holomorphic modular form of weight $4$ which is $E_4$. And, the latter can be expressed as \cite{Handbooks}
\be
  E_4=\theta_2^8-\theta_2^4\theta_3^4+\theta_3^8\,.
\ee
Therefore, in the present case, the MLDE \eqref{eq:2ndmlde} becomes
\be
  \frac{d^2\vec{F}}{d\tau^2}-\frac{i\pi}{3}E_2(\tau)\frac{d\vec{F}}{d\tau}+\alpha_3\pi^2E_4(\tau)\vec{F}=0\,.
\label{eq:2ndmldeO}
\ee

\uwave{In the third step}, in \eqref{eq:2ndmldeO} we substitute the trial solution \eqref{eq:trailsolgen} and the respective $q$-expansions of $E_2(\tau),E_4(\tau)$ as given in appendix \ref{app:qexp}. Solving the indicial equation, we get\footnote{It is worth noting that for real $\alpha_3>-1/36$, we have $s_1<s_2$.}
\bea
  s=s_1,s_2\,,\quad s_1=\frac{1}{6} \left(1-\sqrt{36 \alpha _3+1}\right),\quad s_2=\frac{1}{6} \left(1+\sqrt{36 \alpha _3+1}\right)\,.
\label{eq:svaluesO}
\eea
Let us first take $s=s_1$, and consider the following points on the parameter space.\footnote{The statements that come next are based on studying \eqref{eq:2ndmldeO} at points \eqref{eq:singularptsalpha3} with $m=1,2,\dots,50$.}
\be
  \alpha_3=\frac{1}{36} \left(9 m^2-1\right),\quad m=1,2,3,\dots\,.
\label{eq:singularptsalpha3}
\ee
At any of the above points with even $m$, the coefficients in the $q$-expansion of the l.h.s. of the MLDE \eqref{eq:2ndmldeO} cannot be set to zero consistently with $a_0\neq0$. Therefore, by requiring the existence of a solution with $s=s_1$, we constrain the parameter space as follows.
\be
  \alpha_3\neq\frac{1}{36} \left(9 m^2-1\right),\quad m=2,4,6,\dots\,.
\label{eq:alpha3neq}
\ee
At any of the points \eqref{eq:singularptsalpha3} with odd $m$, \eqref{eq:2ndmldeO} does not fix $a_0,a_m$. In this case, all the $a_r$ with odd $r<m$ are equal to zero, all the $a_r$ with odd $r>m$ are proportional to $a_m$.\footnote{Thus, if we require $a_m$ to be zero, all the odd coefficients vanish.} At a generic point $\alpha_3$ different from \eqref{eq:singularptsalpha3}, only $a_0$ is unfixed. In this case, all the $a_r$ with odd $r$ vanish. In both cases, all the $a_r$ with even $r$ are proportional to $a_0$. The first few are presented below.
{\small
\begin{equation}
\begin{aligned}
  &a_2=-\frac{2 a_0 \left(\sqrt{36 \alpha _3+1}+180 \alpha _3-1\right)}{\sqrt{36 \alpha _3+1}-6}\,, \\
  &a_4=\frac{a_0 \left(7 \left(\sqrt{36 \alpha _3+1}-1\right)+36 \left(25 \sqrt{36 \alpha _3+1}-1800 \alpha _3-129\right) \alpha _3\right)}{18 \sqrt{36 \alpha _3+1}-36 \alpha _3-73}\,, \\
  &a_6=\frac{4 a_0 \left(91 \left(\sqrt{36 \alpha _3+1}-1\right)-6 \alpha _3 \left(180 \alpha _3 \left(1800 \alpha _3-105 \sqrt{36 \alpha _3+1}+958\right)+117 \sqrt{36 \alpha _3+1}+1198\right)\right)}{397 \sqrt{36 \alpha _3+1}+36 \left(\sqrt{36 \alpha _3+1}-36\right) \alpha _3-1332}\,.
\label{eq:coeffaO1}
\end{aligned}
\end{equation}
}%
Analytic expressions of higher even coefficients are more lengthy in terms of $a_0$. However, since we will be using $a_8$ for analytic computations, let us now provide the equation to be solved.
\begin{align}
  &\left(\sqrt{36 \alpha _3+1}-1\right) \left(7 a_0+4 a_2+3 a_4+a_6\right)+2 \left(\sqrt{36 \alpha _3+1}-24\right) a_8 \nn \\
  &\quad+180 \left(73 a_0+28 a_2+9 a_4+a_6\right) \alpha _3-12 \left(4 a_2+6 a_4+3 a_6\right)=0\,.
\label{eq:coeffa8O1}
\end{align}
Let us now take $s=s_2$. In this case, \eqref{eq:2ndmldeO} does not fix $a_0$. All the $a_r$ with odd $r$ vanish, and all the $a_r$ with even $r$ are proportional to $a_0$. The first few even coefficients are presented below.
{\small
\begin{equation}
\begin{aligned}
  &a_2=-\frac{2 a_0 \left(\sqrt{36 \alpha _3+1}-180 \alpha _3+1\right)}{\sqrt{36 \alpha _3+1}+6}\,, \\
  &a_4=\frac{a_0 \left(7 \left(\sqrt{36 \alpha _3+1}+1\right)+36 \left(25 \sqrt{36 \alpha _3+1}+1800 \alpha _3+129\right) \alpha _3\right)}{18 \sqrt{36 \alpha _3+1}+36 \alpha _3+73}\,, \\
  &a_6=\frac{4 a_0 \left(91 \left(\sqrt{36 \alpha _3+1}+1\right)+6  \alpha _3 \left(180 \alpha _3 \left(105 \sqrt{36 \alpha _3+1}+1800 \alpha _3+958\right)-117 \sqrt{36 \alpha _3+1}+1198\right)\right)}{\left(\sqrt{36 \alpha _3+1}+6\right) \left(\sqrt{36 \alpha _3+1}+12\right) \left(\sqrt{36 \alpha _3+1}+18\right)}\,.
\label{eq:coeffaO2}
\end{aligned}
\end{equation}
}%
And, $a_8$ can be obtained by solving the following equation.
\begin{align}
  &\left(\sqrt{36 \alpha _3+1}+1\right) \left(7 a_0+4 a_2+3 a_4+a_6\right)+2 \left(\sqrt{36 \alpha _3+1}+24\right) a_8 \nn \\
  &\quad-180 \left(73 a_0+28 a_2+9 a_4+a_6\right) \alpha _3+12 \left(4 a_2+6 a_4+3 a_6\right)=0\,.
\label{eq:coeffa8O2}
\end{align}

\begin{figure}[h!]
  \centering
    \includegraphics[width=8.0cm]{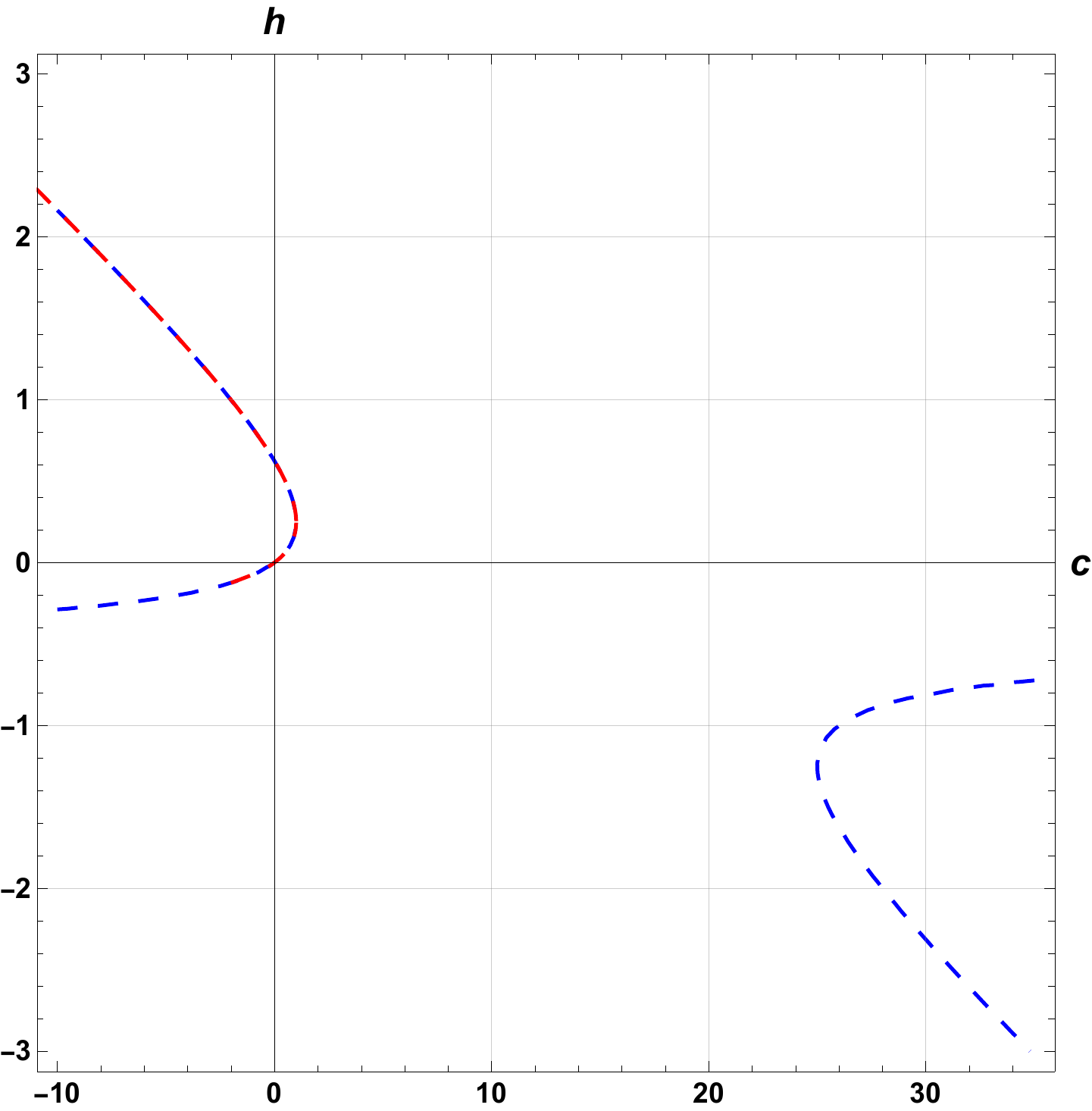}
  \caption{Red dashed line represents the locus of $(c(\alpha_3),h(\alpha_3))$. Blue dashed line represents the locus of the level-2 Virasoro null states.}
\label{fig:chplane}
\end{figure}

\uwave{In the fourth step}, we tune $\alpha_3$ and the above unfixed coefficients so that the 2 solutions found in the previous step are 4-point Virasoro blocks. The relevant CFT quantities are $\{c,h,h_{p(1)},h_{p(2)}\}$.\footnote{Here, $h\equiv h_1=h_2=h_3=h_4,\ \mathfrak{H}=4h$.} We assume identity to be an intermediate operator as compatible with the identical external operators\footnote{The necessary requirements for the occurance of the vacuum block are: $h_1=h_2,\ h_3=h_4$.}, and the dimension of the other intermediate operator to be greater than zero. Thus, in keeping with our convention \eqref{eq:hporder}, we take\footnote{The exclusion of the equality $h_{p(2)}=h_{p(1)}$ is due to the uniqueness of identity operator.}
\be
  h_{p(2)}>h_{p(1)}=0\,.
\ee
By comparing the $q\to0$ limits of the 2 solutions found in the third step with those of \eqref{eq:qexpVblockgen}, we see that the 2 values of $s$ in \eqref{eq:svaluesO} are the candidates for $h_{p(k)}-4h/3,\ k=1,2$. Since $s_1\ngtr s_2$, we equate the first solution i.e. the one with $s=s_1$, with the vacuum block. Hence, $h,h_{p(2)}$ can be expressed in terms of $\alpha_3$ as
\be
  h=\frac{1}{8} \left(\sqrt{36 \alpha _3+1}-1\right),\quad h_{p(2)}=\frac{1}{3} \sqrt{36 \alpha _3+1},\quad \alpha_3>-\frac{1}{36}\,.
\label{eq:halphaO}
\ee
This also fixes $a_0$ of the respective solutions to be $16^s$. For the Virasoro blocks in the present case, all the odd coefficients $f_{2r-1},\ r\geq1$ in \eqref{eq:qexpVblockgen} vanish.\footnote{When $h_1=h_2,\ h_3=h_4$, all the odd coefficients $b_{2r-1},\ r\geq1$ in $H$ identically vanish \cite{CHKL2017}. See appendix \ref{app:Hfunction} for a discussion. \label{fn:vanishoddcoeffVB}} Both the series solutions are automatically tailored to this, except for the first solution we need to fix $a_m=0$ when $\alpha_3=\left(9 m^2-1\right)/36,\ m=1,3,5,\dots$. Now, to express the central charge $c$ in terms of $\alpha_3$, one may equate the ratio $a_2/a_0$ for the first solution to the coefficient $f_2$ for the Virasoro vacuum block:\footnote{Here, we use \eqref{eq:fcoeffsVVB} with $h_L=h_R=h$.}
\begin{align}
  &b_2(c,h,0)+\frac{1-c+32h}{2}=-\frac{2 \left(\sqrt{36 \alpha _3+1}+180 \alpha _3-1\right)}{\sqrt{36 \alpha _3+1}-6}\,. \nn \\
  &\implies c\doteq\frac{7-7 \sqrt{36 \alpha _3+1}-9 \alpha _3 \left(\sqrt{36 \alpha _3+1}-10\right)}{9 \alpha _3-2}\,,
\label{eq:cencO}
\end{align}
where the dotted equality indicates that the limiting value to be taken at $\alpha_3=2/9$. It is straightforward to check that $(c(\alpha_3),h(\alpha_3))$ lies on the locus of the level-2 Virasoro null states:
\be
  16 h^2 +h(2 c-10) +c=0\,.
\ee
However, only a portion of it gets traced when we vary $\alpha_3$, since $c(\alpha_3)\leq1,\ h(\alpha_3)> -1/8$, see Fig. \ref{fig:chplane}.
Therefore, for the case of 4 identical external primaries and 2 intermediate ones, one of which is identity and the other with dimension greater than zero, the MLDE \eqref{eq:2ndmldeO} always corresponds to a second order BPZ equation.\footnote{In the opposite order, second order BPZ equations for identical operators containing a level-2 null state have been brought to the form \eqref{eq:2ndmldeO} in appendix A of \cite{CGL2020}.}

Note that in the present case, the non-vanishing coefficients in the $q$-expansions of the Virasoro blocks are the even ones $f_{2n}$, and they all have $c$-dependencies. The MLDE, on the other hand, computes $a_{2n}/a_0$ in terms of $\alpha_3$ as candidates for them. So far, we have only equated $f_2$ of the vacuum block with $a_2/a_0$ of our first solution, yielding $c(\alpha_3)$ in \eqref{eq:cencO}. We have checked that plugging this $c(\alpha_3)$ into $f_4,f_6,f_8$ of the vacuum block respectively gives $a_4/a_0,a_6/a_0,a_8/a_0$ of the first solution (as in \eqref{eq:coeffaO1},\eqref{eq:coeffa8O1}). We have performed a similar check, relating the coefficients $f_2,f_4,f_6,f_8$ of the excited block (with dimension $h_{p(2)}(\alpha_3)$ as obtained in \eqref{eq:halphaO}) to the coefficients $a_2/a_0,a_4/a_0,a_6/a_0,a_8/a_0$ of our second solution (as in \eqref{eq:coeffaO2},\eqref{eq:coeffa8O2}). For these checks, we have used a Mathematica code published in \cite{CHKL2017} to get the first few analytical coefficients $b_2,b_4,b_6,b_8$ of the $q$-expansion of $H$, and then used its relation to Virasoro blocks \eqref{eq:VBlockH} to generate the required coefficients $f_2,f_4,f_6,f_8$, see appendix \ref{app:Hfunction}. We expect this to hold true for higher coefficients as well, in support of which we present two numerical examples at the end of this subsection.

\uwave{In the fifth step}, we obtain the modular transformation matrices that act on the above solutions to \eqref{eq:2ndmldeO} and then set up equation for the 3-point coefficient $c_{OOp(2)}$ associated with the stripped correlator $G_{OOOO}$. Plugging \eqref{eq:alpharelO} in \eqref{eq:closedsolshol}, and then using \eqref{eq:hypergeoid} we get\footnote{$\Gamma(1–z)\Gamma(z)=\pi\csc\pi z$ has also been used to simplify the expressions.}
\be
\begin{aligned}
  &M(T)=
  \begin{pmatrix}
    e^{\frac{i\pi}{6} \left(1-\sqrt{36 \alpha _3+1}\right)} & 0\\
    0 & e^{\frac{i\pi}{6} \left(1+\sqrt{36 \alpha _3+1}\right)}
  \end{pmatrix}, \\
  &M(S)=
  \begin{pmatrix}
    \frac{1}{2} \csc \left(\frac{\pi}{6}  \sqrt{36 \alpha _3+1}\right) & \frac{\sqrt{\pi } 2^{\frac{1}{3} \sqrt{36 \alpha _3+1}} \csc \left(\frac{\pi}{3}  \sqrt{36 \alpha _3+1}\right) \Gamma \left(1+\frac{1}{6} \sqrt{36 \alpha _3+1}\right)}{\Gamma \left(1-\frac{1}{3} \sqrt{36 \alpha _3+1}\right) \Gamma \left(\frac{1}{2}+\frac{1}{2} \sqrt{36 \alpha _3+1}\right)} \\
    -\frac{\sqrt{\pi } 2^{-\frac{1}{3} \sqrt{36 \alpha _3+1}} \csc \left(\frac{\pi}{3}  \sqrt{36 \alpha _3+1}\right) \Gamma \left(1-\frac{1}{6} \sqrt{36 \alpha _3+1}\right)}{\Gamma \left(\frac{1}{2}-\frac{1}{2} \sqrt{36 \alpha _3+1}\right) \Gamma \left(1+\frac{1}{3} \sqrt{36 \alpha _3+1}\right)} & -\frac{1}{2} \csc \left(\frac{\pi}{6} \sqrt{36 \alpha _3+1}\right)
  \end{pmatrix}.
\end{aligned}
\ee
At this stage, we want to glue the holomorphic blocks with the anti-holomorphic ones. When $-1/36<\alpha_3<20/9,\ \alpha_3\neq2/9$, for each $\alpha_3$ we have an $\bar{\alpha}_3\neq\alpha_3$ with $c(\alpha_3)=c(\bar{\alpha}_3)$, thus, we have possibilities of spinning operators. We take $\bar{\alpha}_3$ to correspond to the anti-holomorphic dimensions, and from aforementioned formulae of dimensions it is clear that the identical external operators and also the excited intermediate operator are all spinning. However, in such cases, the bootstrap condition corresponding to $T$ in \eqref{eq:genbootstrap} alone sets $\mathcal{C}=0$, thereby, the correlator of 4 identical spinning operators vanish. Now, the non-trivial case will be to deal with only scalar operators. In the latter case, with $\alpha_3>-1/36$ the bootstrap condition corresponding to $T$ in \eqref{eq:genbootstrap} allows only vanishing non-diagonal elements in $\mathcal{C}$:
\be
  \mathcal{C}=\begin{pmatrix} 1 & 0 \\ 0 & \beta \end{pmatrix},\quad \beta=c^2_{OOp(2)}\,.
\ee
Now, the $S$-invariance solves for $\beta$ as
\be
  \beta=\frac{2^{\frac{1}{3} (-2) \sqrt{36 \alpha _3+1}} \Gamma \left(-\frac{1}{3} \sqrt{36 \alpha _3+1}\right) \Gamma \left(1-\frac{1}{6} \sqrt{36 \alpha _3+1}\right) \Gamma \left(\frac{1}{2} \sqrt{36 \alpha _3+1}+\frac{1}{2}\right)}{\Gamma \left(\frac{1}{3} \sqrt{36 \alpha _3+1}\right) \Gamma \left(\frac{1}{2}-\frac{1}{2} \sqrt{36 \alpha _3+1}\right) \Gamma \left(\frac{1}{6} \sqrt{36 \alpha _3+1}+1\right)}\,.
\label{eq:3ptsquareO}
\ee

\uwave{In the sixth step}, we constrain the parameter space $\alpha_3$ to meet various unitarity bounds. Note that $\alpha_3> -1/36$ already maintains $h_{p(2)}>0$, see \eqref{eq:halphaO}. $h>0$ now further restricts $\alpha_3>0$. Accordingly, $c$ takes the value $0$ only when $\alpha_3=35/36$, which \eqref{eq:alpha3neq} rules out. As a result, we can only have $c>0$, which limits $0<\alpha_3<35/36$. In fact, in this region of the parameter space, we have $0<c\leq1$, where the equality holds at $\alpha_3=2/9$.

\begin{figure}[h!]
    \centering
    \subfloat[\centering For vaccum pillow block]{{\includegraphics[width=12cm]{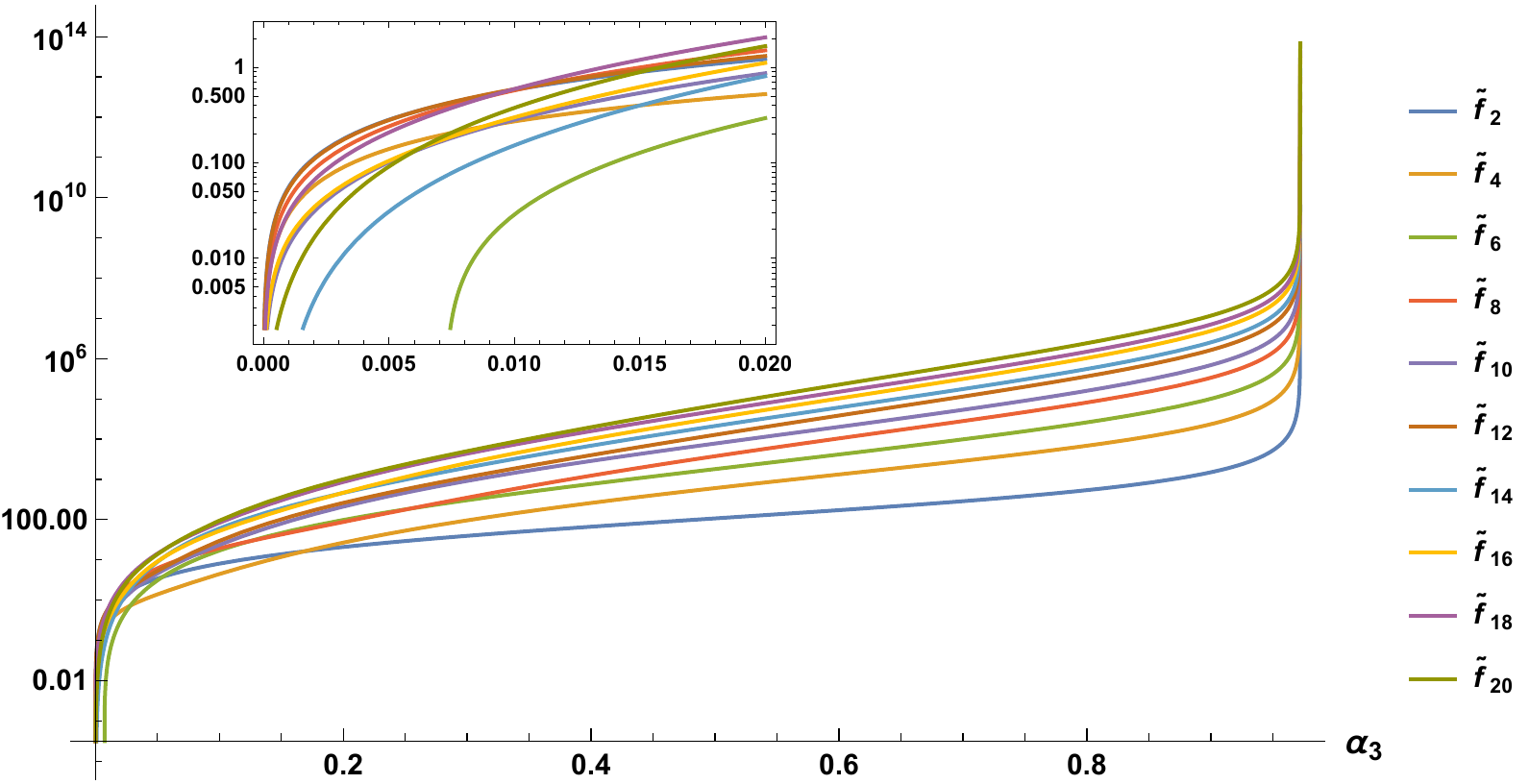} \label{fig:positivityvac}}}
    \qquad
    \subfloat[\centering For excited pillow block]{{\includegraphics[width=12cm]{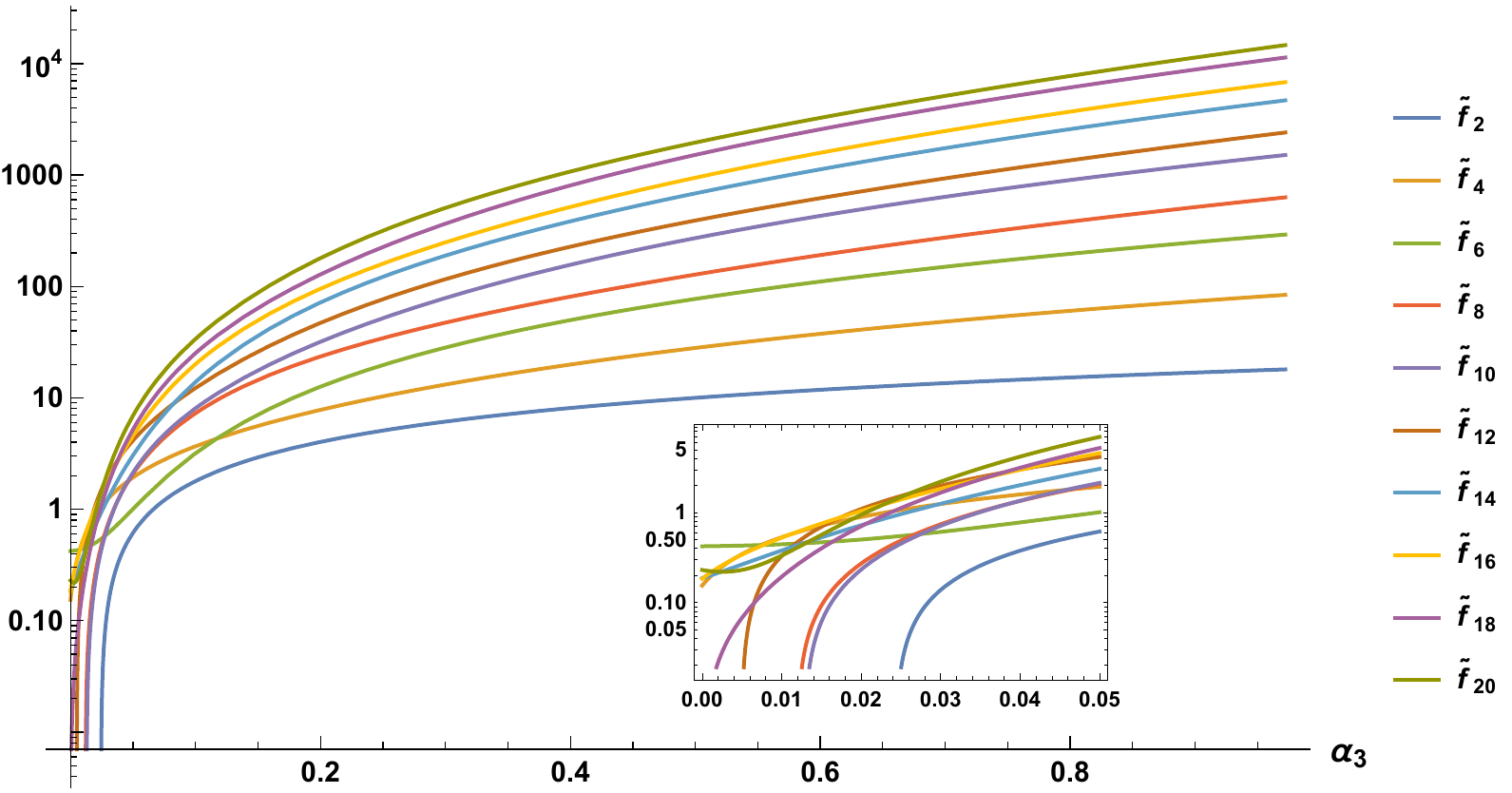} \label{fig:positivityexc}}}
    \caption{First few non-vanishing $q$-expansion coefficients in the pillow blocks are plotted against $\alpha_3$. $y$-axes are taken to be logarithmic.}
    \label{fig:positivitycons}
\end{figure}

We have 4 identical external operators in the present case, thus the non-negativity of the $q$-expansion coefficients of the pillow blocks, as discussed in section \ref{sec:MLDEsph}, applies. Since the odd coefficients $f_{2r+1}$ of the Virasoro blocks vanish, so do the odd coefficients $\tilde{f}_{2n+1}$ of respective pillow blocks in \eqref{eq:genpillowblock}. Now, the non-negativity of the even coefficients $\tilde{f}_{2n}$ of the pillow blocks requires
\be
  \tilde{\gamma}_{n0}+\sum_{r=1}^n\tilde{\gamma}_{nr} \frac{a_{2r}}{a_0}\geq0,\quad n=1,2,3,\dots\,,
\label{eq:unipillowO}
\ee
as per the identification of the Virasoro blocks with the 2 solutions of our MLDE. Corresponding to each solution, by plugging respective $a_{2r}/a_0$ in terms of $\alpha_3$ in \eqref{eq:unipillowO} along with the substitutions \eqref{eq:halphaO} and \eqref{eq:cencO}, we get a countable infinite set of non-negative functions: $\tilde{f}_{2n}(\alpha_3)\geq0,\ n\geq1$. For both the solutions, we analyze the cases with $n=1,\dots,10$. For the first solution i.e. the one identified with the vacuum block, $\tilde{f}_2(\alpha_3),\dots,\tilde{f}_{20}(\alpha_3)$ are non-negative in the region $0<\alpha_3<35/36$, except $\tilde{f}_6(\alpha_3),\tilde{f}_{14}(\alpha_3)$. Now, the requirement of the non-negativity of the last 2 uplifts the lower bound on $\alpha_3$ from $0$ to $\alpha_3^*\approx0.00719982$.\footnote{$\alpha_3^*$ is the smallest positive real root of the polynomial
$$\begin{aligned}
  132860250000000000 x^7-90541248876000000 x^6+67714737664118400 x^5-11791778905265760 x^4\\
  +581776809364116 x^3+1455851807721 x^2-375258931508 x+2439586688\,.
\end{aligned}$$}
In fact, $\alpha_3^*$ comes from the condition $\tilde{f}_{6}(\alpha_3)\geq0$ as the uplift of the lower bound from the condition $\tilde{f}_{14}(\alpha_3)\geq0$ is smaller, see Fig. \ref{fig:positivityvac}. Now, the consideration of the second solution i.e. the one identified with the excited block provides a stronger lower bound on $\alpha_3$. In this case, $\tilde{f}_2(\alpha_3),\dots,\tilde{f}_{20}(\alpha_3)$ are non-negative in the region $0<\alpha_3<35/36$, with the exceptions for $\tilde{f}_2(\alpha_3),\tilde{f}_{8}(\alpha_3),\tilde{f}_{10}(\alpha_3),\tilde{f}_{12}(\alpha_3)$. Now, the requirement of the non-negativity of the last 4 uplifts the lower bound on $\alpha_3$:
\be
  \alpha_3^{**}\leq\alpha_3<\frac{35}{36}\,,
\label{eq:nonnegpil}
\ee
where $\alpha_3^{**}\approx0.0242458$ is the smallest positive real root of the polynomial $2624400 x^3-333720 x^2+3249 x+80$. In fact, $\alpha_3^{**}$ comes from the condition $\tilde{f}_{2}(\alpha_3)\geq0$ as the uplifts of the lower bound from the conditions $\tilde{f}_{8}(\alpha_3),\tilde{f}_{10}(\alpha_3),\tilde{f}_{12}(\alpha_3)\geq0$ are smaller, see Fig. \ref{fig:positivityexc}.

Fig. \ref{fig:positivitycons} shows the semi-logarithmic plots versus $\alpha_3$ of the first few non-vanishing coefficients of the pillow blocks corresponding to the 2 solutions of our MLDE. As can be observed, each of these coefficients is non-negative and continuous on some region of the type $[\hat{\alpha}_3,35/36)$ with $\hat{\alpha}_3<2/9$. If this feature continues to occur in higher coefficients, we may expect further improvement on the lower bound in \eqref{eq:nonnegpil}, but we do not obtain countable points in $\alpha_3$. In contrast to this, we know that for $0<c<1$ the following discrete values are admissible for unitary CFTs \cite{FQS1984,GKO1986}.
\be
  c=1-\frac{6}{l (l+1)},\quad l=3,4,5,\dots\,,
\label{eq:discretecuni}
\ee
which can be attained respectively at\footnote{The lowest such $\alpha_3\ (l=3)$ equals $5/144\approx 0.0347222$, which is well above $\alpha_3^{**}$.}
\be
  \alpha_3=\frac{2}{9}-\frac{l}{(l+1)^2},\ \frac{2}{9}+\frac{l+1}{l^2},\quad  l=3,4,5,\dots\,.
\label{eq:discretealpha3}
\ee
Hence, the above discussion indicates that the non-negativity of the $q$-expansion coefficients of pillow blocks is a necessary but not sufficient condition for unitary CFTs. In support of this, we also present one numerical non-unitary example at the end of this subsection in which the $q$-expansion coefficients of both pillow blocks are tested to be non-negative to order $q^{125}$.

\begin{figure}[h!]
  \centering
    \includegraphics[width=10.0cm]{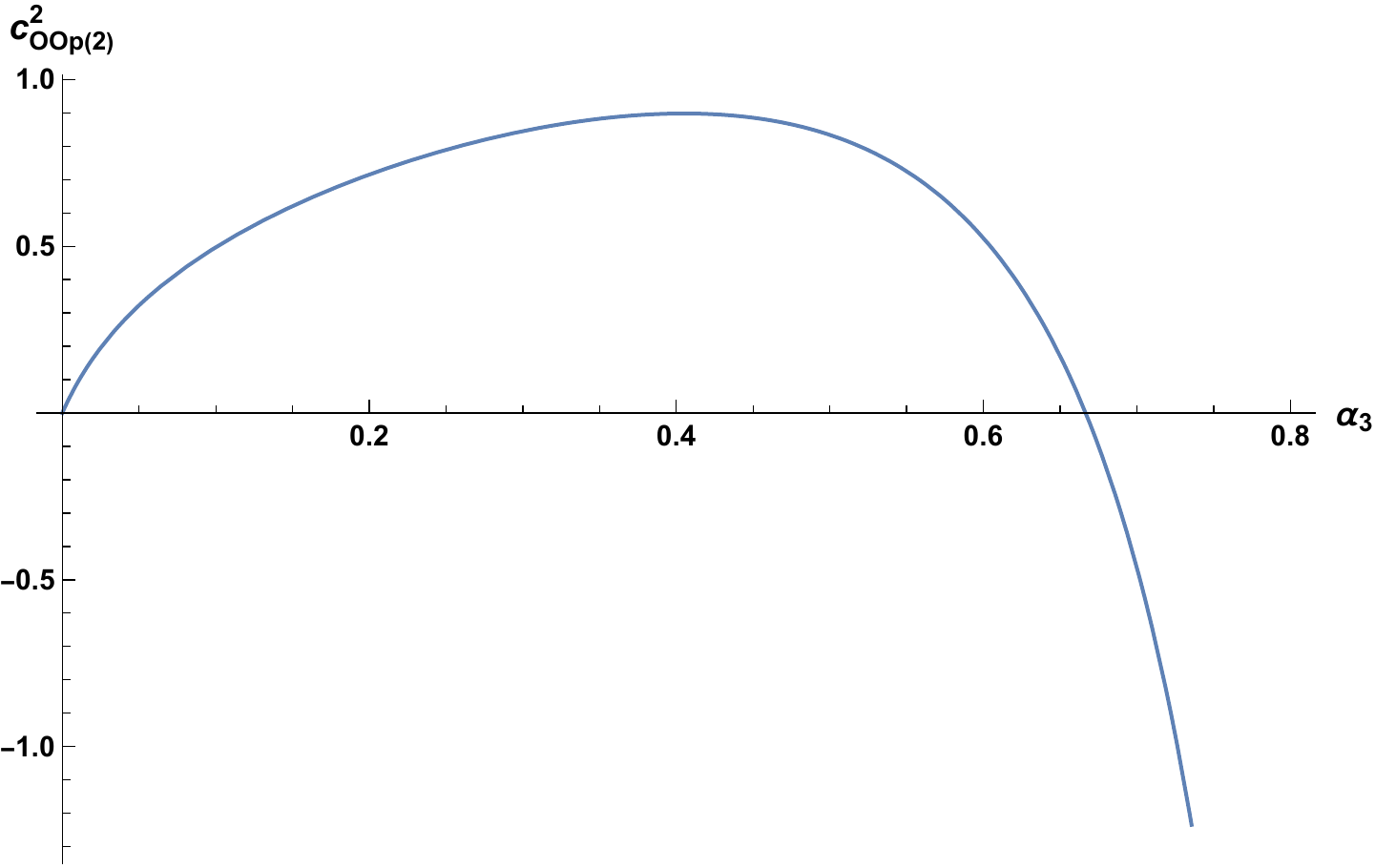}
  \caption{$c^2_{OOp(2)}$ is plotted against $\alpha_3$.}
\label{fig:3ptsquare}
\end{figure}

Now in unitary CFTs, the 3-point coefficient $c_{OOp(2)}$ associated with the stripped correlator $G_{OOOO}$ must be real. Thus, the non-negativity of $c^2_{OOp(2)}$ obtained in \eqref{eq:3ptsquareO} improves the upper bound on $\alpha_3$ as
\be
  \alpha_3^{**}\leq\alpha_3\leq\frac{2}{3}\,,
\ee
see Fig. \ref{fig:3ptsquare}. Note that $2/3$ is also the highest value of $\alpha_3$ allowed in the discrete set \eqref{eq:discretealpha3}.

\subsubsection*{Explicit examples}

\noindent {\underline{Non-unitary}}: Consider the MLDE \eqref{eq:2ndmldeO} with $\alpha_3=1/2$. The associated CFT quantities can be computed as
\be
\begin{aligned}
  &c=\frac{1}{5} \left(104-23 \sqrt{19}\right)\approx 0.749065,\quad h=\frac{1}{8} \left(\sqrt{19}-1\right)\approx 0.419862,\quad h_{p(2)}=\frac{\sqrt{19}}{3}\approx 1.45297\,, \\
  &c^2_{OOp(2)}=\frac{2^{-\frac{2 \sqrt{19}}{3}} \Gamma \left(-\frac{\sqrt{19}}{3}\right) \Gamma \left(1-\frac{\sqrt{19}}{6}\right) \Gamma \left(\frac{1}{2}+\frac{\sqrt{19}}{2}\right)}{\Gamma \left(\frac{\sqrt{19}}{3}\right) \Gamma \left(\frac{1}{2}-\frac{\sqrt{19}}{2}\right) \Gamma \left(1+\frac{\sqrt{19}}{6}\right)}\approx 0.833669\,.
\end{aligned}
\ee
The 2 solutions to the MLDE respectively go as $(16q)^{\frac{1}{6}-\frac{\sqrt{19}}{6}}$ and $(16q)^{\frac{1}{6}+\frac{\sqrt{19}}{6}}$ as $q\to0$. We have solved the MLDE recursion relations numerically and found the $q$-expansion coefficients $a_n$ up to order $q^{125}$ for both solutions, with odd ones vanishing. On the other hand, first using the Mathematica code in \cite{CHKL2017}, the $q$-expansion coefficients $b_n$ of the $H$-functions corresponding to $h_{p(1)}=0$ and $h_{p(2)}=1.45297$ are numerically obtained up to order $q^{125}$, and then the numerical coefficients $f_n$ of the 2 respective Virasoro blocks are calculated coding the relation \eqref{eq:VBHeta}. Indeed, they respectively match the 2 solutions of our MLDE. In the present example, we have found the $q$-expansion coefficients of the blocks to be non-integers.

Knowing above coefficients $a_n$, we have calculated the $q$-expansion coefficients of respective pillow blocks up to order $q^{125}$, with odd ones vanishing. We have found all the even ones \eqref{eq:unipillowO} to be positive and non-linearly growing with $n$ for both the pillow blocks, although the central charge does not lie in \eqref{eq:discretecuni}.

\noindent {\underline{Ising CFT}}: We consider the correlator $\langle\sigma\sigma\sigma\sigma\rangle$ where $\mathds{1}$ and $\varepsilon$ contribute in the intermediate channel. The CFT parameters as well as the Virasoro blocks that contribute to the associated stripped correlator are given by
\be
\begin{aligned}
  &c=\frac{1}{2},\quad h_{\sigma}=\frac{1}{16},\quad h_{\varepsilon}=\frac{1}{2},\quad c^2_{\sigma\sigma\varepsilon}=\frac{1}{4}\,, \\
  &F_{\mathds{1}}(\tau)=\frac{\sqrt{\sqrt{1-\lambda (\tau )}+1}}{\sqrt{2} \sqrt[12]{(1-\lambda (\tau )) \lambda (\tau )}},\quad F_{\varepsilon}(\tau)=\frac{\sqrt{2} \sqrt{1-\sqrt{1-\lambda (\tau )}}}{\sqrt[12]{(1-\lambda (\tau )) \lambda (\tau )}}\,.
\end{aligned}
\ee
Here we record the first few $q$-expansion coefficients:
\begin{align}
  &F_{\mathds{1}}(\tau)=(16q)^{-\frac{1}{12}}\left(1 + 3 q^2 + 4 q^4 + 7 q^6 + 13 q^8 + 19 q^{10}+\cdots\right), \nn \\
  &F_{\varepsilon}(\tau)=(16q)^{\frac{5}{12}}\left(1 + q^2 + 3 q^4 + 4 q^6 + 7 q^8 + 10 q^{10}+\cdots\right).
\end{align}
The above $q$-expansions can be obtained to any desired order $q^n$ employing the $q$-expansion of $\lambda(\tau)$ as given in appendix \ref{app:qexp}. The odd coefficients are zero while all the even ones happen to be positive integers, which we have tested up to order $q^{125}$.

Now considering the MLDE \eqref{eq:2ndmldeO} with $\alpha_3=5/144$, the 2 series solutions generate the aforesaid $q$-expansions, which we have tested up to order $q^{125}$. Also the CFT quantities associated with the MLDE agree with those mentioned above.

\subsection{Pairwise identical operators}
\label{sec:pairwiseid}

Let us take $O_1=O_2\equiv O_L,\ O_3=O_4\equiv O_R$. In this case, $\tilde{\Gamma}_1$ is the subgroup of the modular group generated by $T,ST^2S$. In order for \eqref{eq:2ndmlde} to be an MLDE w.r.t. $\tilde{\Gamma}_1$, we must have
\be
  \alpha_2=\frac{1}{2}-2\alpha_1,\quad \alpha_5=-\alpha_4\,.
\ee
This is due to the following transformations and Jacobi identity \cite{Handbooks}:
\be
  \theta_2(\tau+1)=e^{\frac{i\pi}{4}}\theta_2(\tau),\quad \theta_3(\tau+1)=\theta_4(\tau),\quad \theta_4(\tau)^4=\theta_3(\tau)^4-\theta_2(\tau)^4\,.
\ee
Therefore, in the present case, the MLDE \eqref{eq:2ndmlde} becomes
\begin{align}
  &\frac{d^2\vec{F}}{d\tau^2}-\frac{i\pi}{3}E_2(\tau)\frac{d\vec{F}}{d\tau}+i\pi(\frac{1}{3}-2\alpha_1)\left(\theta_2(\tau)^4-2\theta_3(\tau)^4\right)\frac{d\vec{F}}{d\tau} \nn\\
  &\qquad\qquad\qquad\qquad+\pi^2\left(\alpha_3\theta_2(\tau)^8-\alpha_4\theta_3(\tau)^4\theta_4(\tau)^4\right)\vec{F}=0\,.
\label{eq:2ndmldeLR}
\end{align}
The residual parameter space $\vec{\alpha}=(\alpha_1,\alpha_3,\alpha_4)$ is 3 dimensional.

The series solutions \eqref{eq:trailsolgen} satisfying \eqref{eq:2ndmldeLR} have the following exponents.\footnote{It is worth noting that for real $\alpha_1$ and $\alpha _4<(\frac{1}{2}-2 \alpha _1)^2$, we have $s_1<s_2$.}
\be
  s=s_1,s_2,\quad s_1=\frac{1}{2}-2 \alpha _1-\sqrt{\big(\frac{1}{2}-2 \alpha _1\big)^2-\alpha _4},\quad s_2=\frac{1}{2}-2 \alpha _1+\sqrt{\big(\frac{1}{2}-2 \alpha _1\big)^2-\alpha _4}\,.
\label{eq:expoLR}
\ee
For $s=s_1$, consider the following surfaces.
\be
  \big(\frac{1}{2}-2 \alpha _1\big)^2-\alpha_4=\frac{m^2}{4},\quad m=1,2,3,\dots\,.
\label{eq:singularsurfacalpha14}
\ee
On these surfaces with even $m$, the $q$-expansion coefficients of the l.h.s. of MLDE \eqref{eq:2ndmldeLR} cannot be set to zero consistently with $a_0\neq0$. Therefore, by requiring the existence of a solution with $s=s_1$, we constrain the parameter space as follows.
\be
  \big(\frac{1}{2}-2 \alpha _1\big)^2-\alpha_4\neq\frac{m^2}{4},\quad m=2,4,6,\dots\,.
\label{eq:alpha14neq}
\ee
On the surfaces \eqref{eq:singularsurfacalpha14} with odd $m$, \eqref{eq:2ndmldeLR} does not fix $a_0,a_m$. In this case, all the $a_r$ with odd $r<m$ are equal to zero, all the $a_r$ with odd $r>m$ are proportional to $a_m$.\footnote{Thus, if we require $a_m$ to be zero, all the odd coefficients vanish.} At a generic point $\vec{\alpha}$ away from \eqref{eq:singularsurfacalpha14}, only $a_0$ is unfixed. In this case, all the $a_r$ with odd $r$ vanish. In both cases, all the $a_r$ with even $r$ are proportional to $a_0$. For $s=s_2$, \eqref{eq:2ndmldeLR} does not fix $a_0$. All the $a_r$ with odd $r$ vanish, and all the $a_r$ with even $r$ are proportional to $a_0$.

We will only use $a_2,a_4,a_6,a_8$ for analytic computations. Their analytic expressions in terms of $a_0$ being lengthy, we record here the equations to be solved in the cases $s=s_1,s_2$.
\be
\begin{aligned}
  &a_2 \left(-\alpha _4-\left((s+2) \left(4 \alpha _1+s+1\right)\right)\right)+8 a_0 \left(32 \alpha _3+2 \alpha _4-12 \alpha _1 s+s\right)=0\,, \\
  &a_4 \left(-\alpha _4-\left((s+4) \left(4 \alpha _1+s+3\right)\right)\right)+8 a_2 \left(32 \alpha _3+2 \alpha _4-12 \alpha _1 (s+2)+s+2\right) \\
  &\qquad\qquad\qquad\qquad\qquad-8 a_0 \left(-256 \alpha _3+14 \alpha _4+12 \alpha _1 s+s\right)=0\,, \\
  &a_6 \left(-\alpha _4-s^2-4 \alpha _1 (s+6)-11 s-30\right)+8 a_4 \left(32 \alpha _3+2 \alpha _4-12 \alpha _1 (s+4)+s+4\right) \\
  &-8 a_2 \left(-256 \alpha _3+14 \alpha _4+12 \alpha _1 (s+2)+s+2\right)+32 a_0 \left(224 \alpha _3+14 \alpha _4-12 \alpha _1 s+s\right)=0\,, \\
  &a_8 \left(-\alpha _4-s^2-4 \alpha _1 (s+8)-15 s-56\right)+8 a_6 \left(32 \alpha _3+2 \alpha _4-12 \alpha _1 (s+6)+s+6\right) \\
  &-8 a_4 \left(-256 \alpha _3+14 \alpha _4+12 \alpha _1 (s+4)+s+4\right)+32 a_2 \left(224 \alpha _3+14 \alpha _4-12 \alpha _1 (s+2)+s+2\right) \\
  &\qquad\qquad\qquad\qquad\qquad-8 a_0 \left(-2048 \alpha _3+142 \alpha _4+12 \alpha _1 s+5 s\right)=0\,.
\label{eq:coeffaLR}
\end{aligned}
\ee
The recursion relations for even coefficients involve the lower even coefficients only, which is owing to the fact that $E_2,\theta_2^4-2\theta_3^4,\theta_2^8,\theta_3^4\theta_4^4$ have $q$-expansions with even coefficients.

Now, we tune $\vec{\alpha}$ and the above unfixed coefficients so that the above 2 series solutions are 4-point Virasoro blocks. The associated CFT quantities are $\{c,h_L,h_R,h_{p(1)},h_{p(2)}\}$.\footnote{Here, $h_L\equiv h_1=h_2,\ h_R\equiv h_3=h_4,\ \mathfrak{H}=2(h_L+h_R)$.} We assume identity to be an intermediate operator as compatible with the pairwise identical external operators, and the dimension of the other intermediate operator to be greater than zero. In keeping with our convention \eqref{eq:hporder}, we take
\be
  h_{p(2)}>h_{p(1)}=0\,.
\ee
Since $s_1\ngtr s_2$, the first series solution i.e. the one with $s=s_1$ should be equated with the vacuum block, while the series solution with $s=s_2$ should be equated with the excited block. Comparing the exponents \eqref{eq:expoLR} with those in \eqref{eq:qexpVblockgen}, we get
\begin{align}
  &2(h_L+h_R)=6 \alpha _1-\frac{3}{2}+3 \sqrt{\big(\frac{1}{2}-2 \alpha _1\big)^2-\alpha _4}\,, \label{eq:htotalphaLR}\\
  &h_{p(2)}(\vec{\alpha})=\sqrt{\left(1-4 \alpha _1\right){}^2-4 \alpha _4},\quad \alpha _4<\big(\frac{1}{2}-2 \alpha _1\big)^2,\quad \alpha _1\in \mathbb{R} \label{eq:hp2LR}\,.
\end{align}
This also fixes $a_0$ of the respective solutions to be $16^s$. For the Virasoro blocks in case of pairwise identical external operators, all the odd coefficients $f_{2r-1},\ r\geq1$ in \eqref{eq:qexpVblockgen} vanish.\footnote{See footnote \ref{fn:vanishoddcoeffVB}.} Both the series solutions are automatically tailored to this, except for the first solution we need to fix $a_m=0$ when $\big(\frac{1}{2}-2 \alpha _1\big)^2-\alpha_4=m^2/4,\ m=1,3,5,\dots$. In order to obtain the 3 quantities $c,h_L,h_R$ in terms of $\vec{\alpha}$, one may solve the 3 equations namely \eqref{eq:htotalphaLR}, $f_2=a_2/a_0$ and $f_4=a_4/a_0$, where in each of the last 2, l.h.s. corresponds to the vacuum block \eqref{eq:fcoeffsVVB} while r.h.s. corresponds to the first series solution. However, due to the symmetry of these 3 equations under $h_L\leftrightarrow h_R$, $c(\vec{\alpha}),h_L(\vec{\alpha}),h_R(\vec{\alpha})$ cannot be determined uniquely. In addition, unlike in the case of all identical external operators, the demands to match the higher coefficients in the first solution with respective ones in the vacuum block, as well as the coefficients in the second solution with respective ones in the excited block turn out to be quite restrictive for the parameter space $\vec{\alpha}$ in the present case. Thus, we adopt a different strategy detailed below. For a computer-assisted search for points $\vec{\alpha}$ where both solutions to the MLDE \eqref{eq:2ndmldeLR} are Virasoro blocks, we first simultaneously satisfy the 6 equations in $\{\alpha_1,\alpha_3,\alpha_4,c,h_L,h_R\}$ given below.
\be
\begin{aligned}
  {\text{\it for vacuum block}:}\ \ \ &2(h_L+h_R)-6 \alpha _1-\frac{3}{2}+3 \sqrt{\big(\frac{1}{2}-2 \alpha _1\big)^2-\alpha _4}=0\,, \\
  &f_2\left(c,h_L,h_R,0\right)-\frac{a_2(\vec{\alpha})}{a_0}=0,\quad  f_4\left(c,h_L,h_R,0\right)-\frac{a_4(\vec{\alpha})}{a_0}=0\,, \\
  &f_6\left(c,h_L,h_R,0\right)-\frac{a_6(\vec{\alpha})}{a_0}=0,\quad f_8\left(c,h_L,h_R,0\right)-\frac{a_8(\vec{\alpha})}{a_0}=0\,, \\
  {\text{\it for excited block}:}\ \ \ &f_2\left(c,h_L,h_R,h_{p(2)}(\vec{\alpha})\right)-\frac{a_2(\vec{\alpha})}{a_0}=0\,.
\label{eq:sixLR}
\end{aligned}
\ee
These equations can be written explicitly using \eqref{eq:coeffaLR}, \eqref{eq:hp2LR}, \eqref{eq:fcoeffsVVB}-\eqref{eq:fcoeffsVEB}. They are symmetric under $h_L\leftrightarrow h_R$, involve radicals in $\vec{\alpha}$ and ratios of polynomials with at most the powers $c^6,h_L^4,h_R^4$. We provide a set of explicit points $\{\alpha_1,\alpha_3,\alpha_4,c,h_L,h_R\}$ satisfying above. With these values we numerically check the required validity of \eqref{eq:solVsVB} for higher coefficients in both solutions. In the next step, we use linear perturbations around these points to generate new ones, eventually proving the set to be infinite. The general procedure is as follows. The set of equations \eqref{eq:sixLR} abstractly can be written as
\be
  g^I(t_J)=0,\quad I,J=1,\dots,6\,.
\ee
Suppose a solution $\hat{t}_J$ is a part of a continuous family, then we have
\be
  g^I(\hat{t}_J+\delta t_J)=0\implies \sum_J\partial_Jg^I\Big|_{\hat{t}_J} \delta t_J=0\quad I=1,\dots,6\,,
\ee
with ${\rm rank} (\partial_Jg^I\big|_{\hat{t}_J})\leq5$. Solving above equations allows us to choose some of the $\delta t_J$, the remaining ones are linearly related to them. In the case ${\rm rank} (\partial_Jg^I\big|_{\hat{t}_J})=6$, we have trivial solution: $\delta t_J=0\ \forall J$, and the solution $\hat{t}_J$ is likely to be an isolated point, unless non-linear perturbations come into play.

Let us present here 2 points satisfying \eqref{eq:sixLR}. Subsequently we discuss linear perturbations around them in detail.
\begin{align}
  &\alpha_1=-\frac{1}{3},\quad \alpha_3=-\frac{13}{198},\quad \alpha_4=-3 \sqrt{\frac{2}{11}}-\frac{52}{99},\quad c=82 \sqrt{\frac{2}{11}}-34\approx 0.964917\,, \nn \\
  &h_L=\frac{5}{\sqrt{22}}-1\approx 0.0660036,\quad h_R=2 \sqrt{\frac{2}{11}}\approx 0.852803\,, \label{eq:nonBPZex1LR}\\
  &h_{p(2)}=6 \sqrt{\frac{2}{11}}+1\approx 3.55841\,; \nn \\
  &\alpha_1=\frac{1}{2},\quad \alpha_3=1,\quad \alpha_4=0,\quad c=1,\quad h_L=\frac{1}{4},\quad h_R=\frac{5}{4},\quad h_{p(2)}=1\,. \label{eq:BPZex1LR}
\end{align}
\underline{Non-BPZ cases}: For the example \eqref{eq:nonBPZex1LR}, with the values of $\alpha_1,\alpha_3,\alpha_4$ the 2 solutions to the MLDE \eqref{eq:2ndmldeLR} respectively go as $(16q)^{\frac{2}{3}-3 \sqrt{\frac{2}{11}}}$ and $(16q)^{3 \sqrt{\frac{2}{11}}+\frac{5}{3}}$ as $q\to0$. We have solved the MLDE recursion relations numerically and found the $q$-expansion coefficients $a_n$ up to order $q^{125}$ for both solutions, with odd ones vanishing. On the other hand with the values of $c,h_L,h_R$, first using the Mathematica code in \cite{CHKL2017}, the $q$-expansion coefficients $b_n$ of the $H$-functions corresponding to $h_{p(1)}=0$ and $h_{p(2)}=3.55841$ are numerically obtained up to order $q^{125}$, and then the numerical coefficients $f_n$ of the 2 respective Virasoro blocks are calculated coding the relation \eqref{eq:VBHeta}. Indeed, they respectively match to the 2 solutions of our MLDE. In the present example, we have found the $q$-expansion coefficients of the blocks to be non-integers.

Note that none of the $(c,h_L),(c,h_R)$ lies on the locus of the level-2 Virasoro null states:
\be
  16 h_L^2 +h_L(2 c-10) +c\neq0,\quad 16 h_R^2 +h_R(2 c-10) +c\neq0\,.
\label{eq:awayNull}
\ee
Hence, corresponding to $\vec{\alpha}$ in \eqref{eq:nonBPZex1LR}, we have a second order MLDE \eqref{eq:2ndmldeLR} whose solutions are 4-point Virasoro blocks and the MLDE does not correspond to a BPZ equation.

At \eqref{eq:nonBPZex1LR}, ${\rm rank} (\partial_Jg^I)=5$ and we have
\be
\begin{aligned}
  &\delta \alpha _3= \frac{1}{726} (385-18 \sqrt{22}) \delta \alpha _1,\quad \delta \alpha _4= \frac{1}{726} (945 \sqrt{22}+1364) \delta \alpha _1,\\
  &\delta c= \frac{3}{11} (188-41 \sqrt{22}) \delta \alpha _1,\quad \delta h_L= -\frac{3}{44} (5 \sqrt{22}-24) \delta \alpha _1,\\
  &\delta h_R= -\frac{3}{11} (\sqrt{22}+4) \delta \alpha _1\,.
\label{eq:nonBPZcurve}
\end{aligned}
\ee
Knowing the above first derivatives w.r.t. $\alpha_1$ at one point, one cannot trace the full curve away from \eqref{eq:nonBPZex1LR}, however, guided by the above formulae nearby points satisfying \eqref{eq:sixLR} can be scanned. Here we present one new point near \eqref{eq:nonBPZex1LR} well approximated by \eqref{eq:nonBPZcurve} with $\delta\alpha_1=10^{-4}$, and indeed the errors are of the order $(\delta\alpha_1)^2$.
\be
\begin{aligned}
  &\alpha_1=-\frac{9997}{30000},\quad \alpha_3=\frac{-54 \sqrt{549835009}-7144225927}{108900000000}\,, \\
  &\alpha_4=\frac{-1484514 \sqrt{549835009}-14294885369}{27225000000},\quad c=\frac{41 \sqrt{549835009}-934859}{27500}\,, \\
  &h_L=\frac{5 \sqrt{549835009}-109982}{110000},\quad h_R=\frac{\sqrt{549835009}-3}{27500},\quad h_{p(2)}=\frac{3 \sqrt{549835009}+27491}{27500}\,.
\label{eq:nonBPZex2LR}
\end{aligned}
\ee
Moreover, with \eqref{eq:nonBPZex2LR} we have numerically tested the required validity of \eqref{eq:solVsVB} for higher coefficients in both solutions up to order $q^{125}$, in a similar way to the previous case. We have found the $q$-expansion coefficients of the solutions to be non-integers. In this case as well, the MLDE is a non-BPZ one since \eqref{eq:awayNull} holds.

\noindent\underline{BPZ cases}: For the example \eqref{eq:BPZex1LR}, with the values of $\alpha_1,\alpha_3,\alpha_4$ the 2 solutions to the MLDE \eqref{eq:2ndmldeLR} respectively go as $(16q)^{-1}$ and $(16q)^{0}$ as $q\to0$. We have solved the MLDE recursion relations and found the $q$-expansion coefficients $a_n$ up to order $q^{21}$ for both solutions, with odd ones vanishing.\footnote{For the first solution, we set $a_1=0$ by hand as $\vec{\alpha}$ lies on the surface \eqref{eq:singularsurfacalpha14} with $m=1$.} On the other hand with the values of $c,h_L,h_R$, first using the Mathematica code in \cite{CHKL2017}, the $q$-expansion coefficients $b_n$ of the $H$-functions corresponding to $h_{p(1)}=0$ and $h_{p(2)}=1$ are obtained up to order $q^{21}$, and then the coefficients $f_n$ of the 2 respective Virasoro blocks are calculated coding the relation \eqref{eq:VBHeta}. Indeed, they respectively match to the 2 solutions of our MLDE. In the present example, we have found the $q$-expansion coefficients of the blocks to be non-integers. The full computation has been done analytically.\footnote{For $c=1$ i.e. $b=i$ [see, definition \eqref{eq:chbparameter}], the numerical code in \cite{CHKL2017} breaks. Hence, taking the external dimensions $h_L=1/4,h_R=5/4$, we use the code for the analytical computations of the first 21 $q$-expansion coefficients in the $H$-functions corresponding to 2 Virasoro blocks with above intermediate dimensions and arbitrary $c$. Then we take $b\to i$ limit.}

Clearly $(c,h_L)$ lies on the locus of the level-2 Virasoro null states:
\be
  16 h_L^2 +h_L(2 c-10) +c=0,\quad 16 h_R^2 +h_R(2 c-10) +c\neq0\,.
\label{eq:semiNull}
\ee
Hence, the MLDE corresponds to a BPZ equation in the present case.

At \eqref{eq:BPZex1LR}, ${\rm rank} (\partial_Jg^I)=4$ and we have
\be
\begin{aligned}
  \delta \alpha _4= 3 \delta \alpha _1-\frac{\delta \alpha _3}{2},\quad \delta h_L= \frac{3}{8} (\delta \alpha _3-2 \delta \alpha _1),\quad \delta h_R= \frac{3}{8} (6 \delta \alpha _1+\delta \alpha _3),\quad \delta c= 0\,.
\label{eq:BPZsurface}
\end{aligned}
\ee
This is a 2 dimensional space. The points on it are not guaranteed to correspond BPZ cases. In addition, there could be higher order perturbations that may alter $c$ from 1. However, the curve satisfying \eqref{eq:sixLR} and passing trough \eqref{eq:BPZex1LR} such that always $c=1$ and the MLDE corresponds to a BPZ equation i.e. $h_L=1/4$ can be easily obtained as
\be
  \alpha_3(\alpha_1)=-\alpha _1\left(\alpha _1-3\right)-\frac{1}{4},\quad \alpha_4(\alpha_1)=2 \alpha _1 \left(2 \alpha _1-1\right),\quad h_R(\alpha_1)=3 \alpha _1-\frac{1}{4}\,.
\label{eq:BPZcurvealphaLR}
\ee
Along the curve, $h_{p(2)}=1$. And, the 2 solutions to the MLDE \eqref{eq:2ndmldeLR} respectively go as $(16q)^{-2\alpha_1}$ and $(16q)^{1-2\alpha_1}$ as $q\to0$. As the above $\vec{\alpha}$ lies on the surface \eqref{eq:singularsurfacalpha14} with $m=1$,  we set $a_1=0$ by hand for the first solution. With this, the odd coefficients in both solutions vanish as required for the 2 Virasoro blocks corresponding to $h_{p(1)}=0$ and $h_{p(2)}=1$. We have analytically tested the required validity of \eqref{eq:solVsVB} in both solutions up to order $q^{21}$.

Now, we present one new point near \eqref{eq:BPZex1LR} for which $c\neq1$, nonetheless well approximated by \eqref{eq:BPZsurface} with $\delta\alpha_1=10^{-4},\delta\alpha_3=10^{-4}$ up to errors of order $(\delta\alpha_1)^2,(\delta\alpha_3)^2,\delta\alpha_1\delta\alpha_3$.
\be
\begin{aligned}
  &\alpha_1=\frac{5001}{10000},\quad \alpha_3=\frac{10001}{10000},\quad \alpha_4=\frac{10003 \sqrt{400080013}-200030003}{200000000}\,, \\
  &c=\frac{13002600310009-600060003 \sqrt{400080013}}{1000200010000},\quad h_L=\frac{3 \sqrt{400080013}-40009}{80000}\,, \\
  &h_R=\frac{3 \sqrt{400080013}+40015}{80000},\quad h_{p(2)}=\frac{\sqrt{400080013}-10003}{10000}\,.
\label{eq:BPZex2LR}
\end{aligned}
\ee
Moreover, with \eqref{eq:BPZex2LR} we have numerically tested the required validity of \eqref{eq:solVsVB} for higher coefficients in both solutions up to order $q^{125}$. We have found the $q$-expansion coefficients of the solutions to be non-integers. In this case as well, the MLDE is a BPZ one since \eqref{eq:semiNull} holds.

Now, the modular transformation matrices that act on the above solutions to \eqref{eq:2ndmldeLR} are\footnote{Computed using \eqref{eq:actionSTTSonF}.}
\be
\begin{aligned}
  &M(T)=
  \begin{pmatrix}
    e^{\frac{ i \pi  }{2}\left(1-4 \alpha _1-\sqrt{\kappa_1}\right)} & 0\\
    0 & e^{\frac{ i \pi  }{2}\left(1-4 \alpha _1+\sqrt{\kappa_1}\right)}
  \end{pmatrix},
\end{aligned}
\ee
\begin{align}
  &M(ST^2S)_{(1,1)}=-i e^{\frac{1}{2} i \pi  \left(4 \alpha _1+\sqrt{\kappa _1}\right)} \cos \left(2 \pi  \sqrt{\kappa _2}\right) \csc \big(\frac{\pi  \sqrt{\kappa _1}}{2}\big)\,, \nn \\
  &M(ST^2S)_{(2,1)}=-\frac{2^{-2 \sqrt{\kappa _1}-1} \left((-1)^{4 \sqrt{\kappa _2}}-1\right) (-1)^{2 \alpha _1-2 \sqrt{\kappa _2}} \csc \left(2 \pi  \sqrt{\kappa _2}\right) \Gamma \big(1-\frac{\sqrt{\kappa _1}}{2}\big) \Gamma \big(-\frac{\sqrt{\kappa _1}}{2}\big)}{\Gamma \big(-\frac{\sqrt{\kappa _1}}{2}-2 \sqrt{\kappa _2}+\frac{1}{2}\big) \Gamma \big(-\frac{\sqrt{\kappa _1}}{2}+2 \sqrt{\kappa _2}+\frac{1}{2}\big)}\,, \nn \\
  &M(ST^2S)_{(1,2)}=-\frac{2 i \pi  e^{2 i \pi  \alpha _1} \Gamma \big(\sqrt{\kappa _1}+1\big) \Gamma \big(\sqrt{\kappa _1}\big)}{{\Gamma \big(\frac{1}{2} (\sqrt{\kappa _1}+1)\big)}^2\ \Gamma \big(\frac{1}{2} (\sqrt{\kappa _1}-4 \sqrt{\kappa _2}+1)\big) \Gamma \big(\frac{1}{2} (\sqrt{\kappa _1}+4 \sqrt{\kappa _2}+1)\big)}\,, \nn \\
  &M(ST^2S)_{(2,2)}=i e^{-\frac{1}{2} i \pi  \left(\sqrt{\kappa _1}-4 \alpha _1\right)} \cos \left(2 \pi  \sqrt{\kappa _2}\right) \csc \big(\frac{\pi  \sqrt{\kappa _1}}{2}\big)\,, \nn \\
  &\kappa_1=\left(1-4 \alpha _1\right){}^2-4 \alpha _4,\quad \kappa_2=\alpha _1^2+\alpha _3\,.
\end{align}
Now confining us to scalar operators, the bootstrap condition corresponding to $T$ in \eqref{eq:genbootstrap} allows only vanishing non-diagonal elements in $\mathcal{C}$:
\be
  \mathcal{C}=\begin{pmatrix} 1 & 0 \\ 0 & \beta \end{pmatrix},\quad \beta=c_{LLp(2)}\cdot c_{RRp(2)}\,,
\ee
where $c_{LLp(2)},c_{RRp(2)}$ are the 3-point coefficients associated with the stripped correlator $G_{LLRR}$. Now, the $S$-invariance solves for $\beta$ as\footnote{Here, $^*$ denotes the complex conjugate. We know that $\kappa_1>0$ due to \eqref{eq:hp2LR}.}
\be
\begin{aligned}
  \beta=&-\frac{{\Gamma \big(\frac{1}{2} (\sqrt{\kappa _1}+1)\big)}^4 \left(\cos \big(2 \pi  \sqrt{\kappa _2}\big) \csc ^2\big(\frac{\pi  \sqrt{\kappa _1}}{2}\big) \cos \big(2 \pi  {\sqrt{\kappa _2}}^*\big)-1\right)}{4 \pi ^2 {\Gamma \big(\sqrt{\kappa _1}+1\big)}^2\ {\Gamma \big(\sqrt{\kappa _1}\big)}^2}\\
  &\times \Gamma \big(\frac{1}{2} (\sqrt{\kappa _1}-4 \sqrt{\kappa _2}+1)\big) \Gamma \big(\frac{1}{2} (\sqrt{\kappa _1}+4 \sqrt{\kappa _2}+1)\big) \Gamma \big(\frac{1}{2} (\sqrt{\kappa _1}-4 {\sqrt{\kappa _2}}^*+1)\big) \Gamma \big(\frac{1}{2} (4 {\sqrt{\kappa _2}}^*+\sqrt{\kappa _1}+1)\big)\,.
\label{eq:3ptsprodLR}
\end{aligned}
\ee
Thus, in the aforesaid explicit examples we have
\be
\begin{aligned}
  {\text{\it for \eqref{eq:nonBPZex1LR}}},\quad \beta&=-\frac{16^{-6 \sqrt{\frac{2}{11}}-1} {\Gamma \big(-3 \sqrt{\frac{2}{11}}-\frac{1}{2}\big)}^2 {\Gamma \big(\frac{1}{2}-3 \sqrt{\frac{2}{11}}\big)}^2}{\bigg(\frac{1}{\left(1-2 \cos \left(2 \sqrt{\frac{2}{11}} \pi \right)\right)^2}-1\bigg) {\Gamma \big(-4 \sqrt{\frac{2}{11}}\big)}^2 {\Gamma \big(-2 \sqrt{\frac{2}{11}}\big)}^2}\\
  &\approx 0.0000419138\,; \\
  {\text{\it for \eqref{eq:BPZex1LR}}},\quad \beta&=\frac{5}{4}\,; \\
  {\text{\it for \eqref{eq:nonBPZex2LR}}},\quad \beta&\approx 0.0000419882\,; \\
  {\text{\it for \eqref{eq:BPZex2LR}}},\quad \beta&\approx 1.24968\,.
\end{aligned}
\ee
And, along the curve \eqref{eq:BPZcurvealphaLR}
\be
  \beta=\frac{1}{4} \left| 1-12 \alpha _1\right| \cot \big(\pi  \sqrt{12 \alpha _1-1}\big) \tan \big(\pi  {\sqrt{12 \alpha _1-1}}^*\big)\,.
\ee

Note that with the pairing $LLRR$ of external operators the non-negativity of the $q$-expansion coefficients of pillow blocks does not apply as unitarity bound, see section \ref{sec:MLDEsph}. In all the examples above, $c,h_L,h_R,h_{p(2)}$ are greater than zero, except in the case \eqref{eq:BPZcurvealphaLR} we need to restrict $\alpha_1>1/12$. However, the non-BPZ examples \eqref{eq:nonBPZex1LR}, \eqref{eq:nonBPZex2LR} and the BPZ example \eqref{eq:BPZex2LR} are non-unitary since the respective central charges are less than 1 but do not lie in \eqref{eq:discretecuni}. On the other hand, in the BPZ example \eqref{eq:BPZex1LR} with $c=1$, the representations with respective $h_L,h_R,h_{p(2)}$ are all unitary.\footnote{The Kac determinants for a representation with $(c=1,h>0)$ are non-negative, and respective Gram matrices are positive semi-definite \cite{Francesco1997}.}

\section{Discussions}
\label{sec:discuss}

In the above, we have constructed MLDEs w.r.t. subgroups of modular group whose solutions are Virasoro conformal blocks that appear in a crossing symmetric 4-point correlator on the sphere. We have particularly focused on second order MLDEs with holomorphic coefficients in the cases of all identical and pairwise identical ($LLRR$) external operators, namely \eqref{eq:2ndmldeO} and \eqref{eq:2ndmldeLR}. In the first case, the relevant subgroup is the full modular group, and in the second case, it is generated by $T,ST^2S$. The central charge, operator dimensions of both external as well intermediate ones are functions of parameters that appeared in the MLDEs. We are able to provide these functional dependencies explicitly in the first case and in a subcase of the second case, elsewhere they are implicit. This has been done by comparing the $q$-expansions of the solutions to the MLDEs with those of Virasoro blocks. Using actions of respective subgroups, bootstrap equations involving the associated 3-point coefficients have been set up and solved as well in terms of the MLDE parameters. For the case of all identical external operators, our MLDE with one single parameter corresponds to BPZ equation. We have illustrated one unitary example and one non-unitary example at the end of section \ref{sec:allidenO}. The non-negativity of the $q$-expansion coefficients in the pillow blocks applies in this case as unitarity bound, and we have shown this to be an insufficient condition for unitarity by calculating the coefficients in pillow blocks to be non-negative for our non-unitary example as well. For the case of pairwise identical operators, starting from a few specific examples of MLDEs we have used linear perturbations around them to generate (infinite-)family of MLDEs in section \ref{sec:pairwiseid}. In this case, some of the MLDEs presented do not correspond to BPZ equations. These are novel second order differential equations whose solutions are Virasoro blocks. This itself is quite exciting result in the context of  
building differential equations for the blocks. Also, out of those blocks crossing symmetric correlators can be written as our bootstrap equations possess nontrivial solutions in those cases. In these specific examples, from the corresponding central charges it is clear that the respective representations are non-unitary. With the classification of non-unitary CFTs with $c<1$, one will be able to identify the respective CFTs where our $(c,h)$-values lie.

Below, we discuss potential generalisations of the present analysis and related future avenues, including the consideration of meromorphic forms as the coefficients in our MLDEs in some detail.

\noindent\underline{\it 1. Meromorphic forms}: In the above, as the coefficients of our MLDEs we have considered holomorphic forms, see \eqref{eq:2ndmlde}. The ansatz \eqref{eq:2ndmlde} leads to the most general second order linear differential equation with 3 regular singular points in variable $x=\lambda(\tau)$ namely at $x=0,1,\infty$. Thus, considerations of meromorphic $\phi_0,\phi_1$ should result in more and/or new type singular points of the differential equation in $x$. However, here we point out that such considerations will be important for writing down MLDEs for 4-point Virasoro blocks. Consider the following MLDE w.r.t. the full modular group.
\be
\begin{aligned}
  &\frac{d^2\vec{F}}{d\tau^2}-\frac{ i \pi}{3}  E_2(\tau ) \frac{d\vec{F}}{d\tau}+2 i \pi  \alpha _1\frac{E_6(\tau )}{E_4(\tau )} \frac{d\vec{F}}{d\tau}+\pi ^2 \left(\alpha _3\frac{ E_6(\tau )^2}{E_4(\tau )^2}+\alpha _2 E_4(\tau )\right) \vec{F}=0\,, \\
  &\alpha_1= \frac{2}{3},\quad \alpha_2= \frac{164}{225},\quad \alpha_3= -\frac{8}{9}\,.
\label{eq:2ndmldemeroO}
\end{aligned}
\ee
$E_4$ has zeros at $\tau=e^{\frac{i \pi }{3}},e^{\frac{2i \pi }{3}}$ which correspond to $x=(-1)^{1/3},-(-1)^{2/3}$ respectively. $E_6$ is non-vanishing at these points. The 2 solutions to the MLDE \eqref{eq:2ndmldemeroO} respectively go as $(16q)^{-\frac{4}{5}}$ and $(16q)^{-\frac{1}{5}}$ as $q\to0$. Now considering 4 identical external operators of dimension $h=3/5$ in a CFT with central charge $c=7/10$, the above solutions respectively coincide with the Virasoro blocks corresponding to the intermediate dimensions $h_{p(1)}=0$ and $h_{p(2)}=3/5$. We have analytically tested the required validity of \eqref{eq:solVsVB} for $q$-expansion coefficients in both solutions up to order $q^{15}$. In doing so, the $q$-expansion coefficients in the solutions are obtained solving the above MLDE recursively. For $c=7/10$ i.e. $b=i\sqrt{5}/2$ [see, definition \eqref{eq:chbparameter}], the numerical code in \cite{CHKL2017} breaks. Thus, taking external dimensions $h=3/5$, we use the code for the analytical computations of the first 15 $q$-expansion coefficients in the $H$-functions corresponding to 2 Virasoro blocks with above intermediate dimensions and arbitrary $c$. Then we take $b\to i\sqrt{5}/2$ limit. These coefficients are then used to calculate the coefficients in the Virasoro blocks coding the relation \eqref{eq:VBHeta}. In the present example, we have found the $q$-expansion coefficients of the solutions to be non-integers. The MLDE \eqref{eq:2ndmldemeroO} is a non-BPZ one as $16 h^2 +h(2 c-10) +c\neq0$.

Once the above solutions are identified with the Virasoro blocks, one can solve the MLDE recursively to obtain their $q$-expansions to any desired order $q^n$. We have done this numerically and found the coefficients up to order $q^{100}$ within an accuracy $10^{-50}$. Within this accuracy, the two solutions vanishes at $q=i e^{-\sqrt{3} \pi/2},-i e^{-\sqrt{3} \pi/2}$ i.e. respectively at $\tau=e^{\frac{i \pi }{3}},e^{\frac{2i \pi }{3}}$. Thus, the solutions are analytic although the MLDE has singularities at these points. In the variable $x=\lambda(\tau)$, \eqref{eq:2ndmldemeroO} becomes
\be
\begin{aligned}
  &\frac{d^2\vec{F}}{dx^2}+P(x)\frac{d\vec{F}}{dx}+Q(x)\vec{F}=0,\quad P(x)=\frac{2 (2 x-1)}{ x (x-1)\left(x^2-x+1\right)}\,, \\
  &Q(x)=\frac{2 \left(2 x^6-6 x^5-63 x^4+136 x^3-63 x^2-6 x+2\right)}{25 x^2 (x-1)^2 \left(x^2-x+1\right)^2}\,.
\label{eq:2ndodemeroOx}
\end{aligned}
\ee
Clearly $x=(-1)^{1/3},-(-1)^{2/3}$ are regular singular points of the above ODE. Nonetheless, at these points the 2 general solutions given below are analytic with vanishing constant term in respective Taylor series.
\be
  \frac{((x-1) x+1) P_{-\frac{1}{5}}^{\frac{3}{5}}(2 x-1)}{\sqrt{-((x-1) x)}},\quad \frac{((x-1) x+1) Q_{-\frac{1}{5}}^{\frac{3}{5}}(2 x-1)}{\sqrt{-((x-1) x)}}\,,
\label{eq:solxLegendre}
\ee
where $P^m_n(x)$ denotes the associated Legendre polynomial and $Q^m_n(x)$ the associated Legendre function of the second kind \cite{MathWorld}. As the 2 aforesaid Virasoro blocks can be written as linear combinations of \eqref{eq:solxLegendre}, they are analytic at $x=(-1)^{1/3},-(-1)^{2/3}$ as well, and vanish at these points.

The above discussion shows the importance to scan over MLDEs with meromorphic coefficients in order to construct differential equations for 4-point Virasoro blocks. Such an MLDE may lead to new singularities as well in the Virasoro blocks in addition to the expected ones, unlike the above example.\footnote{For $W_3$ conformal blocks, such a situation has been reported earlier in \cite{FR2011}.} Such a case will be interesting as this can not be observed from the $q$-expansions of the Virasoro blocks due to Zamolodchikov. We leave the systematic study for the future.

\noindent\underline{\it 2. MLDEs with $c>1,\ h_a,h_{p(k)}>0$}: The examples constructed in this paper are with $c\leq1$. It will be important to scan the parameter space $\vec{\alpha}$ for points such that the MLDE \eqref{eq:genmlde} gives Virasoro blocks with $c>1,\ h_a,h_{p(k)}>0$ when $O_{p(k)}\neq\mathds{1}$. If it exists, the MLDE will be a non-BPZ one, additionally the corresponding CFT will be unitary as all representations with $(c>1,h>0)$ are unitary. Such a CFT contains infinite number of Virasoro primaries in its spectrum \cite{Cardy1986}. Thus, above MLDE having a finite order $N$ would reveal non-trivial information about the fusion rules, such as \eqref{eq:genfusionN} or a truncation.

To initiate, let us take the case of pairwise identical operators studied in this paper. In this case, existence of an MLDE \eqref{eq:2ndmldeLR} that corresponds to $c>1,\ h_L,h_R,h_{p(2)}>0\ (h_{p(1)}=0)$ is not ruled out yet, unlike the case of all identical operators. We have performed a preliminary search as follows. Here, we deal with an inverse problem of the one discussed in section \ref{sec:pairwiseid}. We use the first 3 equations in \eqref{eq:sixLR} to express $\alpha_1,\alpha_3,\alpha_4$ in terms of $c,h_L,h_R$, then try to satisfy the remaining 3 equations in the region $c>1,\ h_L,h_R>0$ and single out the cases with $h_{p(2)}>0$. For this, we numerically vary $c$ from $1.025$ to $3$ with step-size $0.025$, and each $h_L,h_R$ from $0.025$ to $2$ with step-size $0.025$. This generates $259200$ points in the above region. For each of these, we numerically solve the first 3 equations in \eqref{eq:sixLR} to get $\alpha_1,\alpha_3,\alpha_4$ and with these values we check the validity of the remaining 3 equations within a tolerance $0.1$. We have taken a large tolerance, keeping in mind that the desired point satisfying \eqref{eq:sixLR} can possibly be an isolated one and may lie only near any of the generated points. For none of the $259200$ points generated, \eqref{eq:sixLR} can be satisfied within above tolerance. Increasing the upper limit of $c,h_L,h_R$ and/or decreasing the step-sizes make the numerics more involved, which we leave for the future.

\noindent\underline{\it 3. Sphere-torus correspondence}: Recently \cite{CGL2020} reported a new sphere-torus correspondence that relates the characters of diagonal theories to specific 4-point conformal blocks on the sphere. Their proposed correspondence considers on the torus-side a CFT denoted by $\mathcal{T}_t$ and on the sphere-side a CFT given by $\mathcal{T}_t \times \mathcal{T}_t /\mathbb{Z}_2$. In the present context of MLDEs, a potential future direction will be to classify all MLDEs (second or higher order, with holomorphic and meromorphic coefficients as well) in the case of all identical operators (as here the relevant subgroup is the full modular group as evident from Tab. \ref{tab:stabilizer}) whose solutions are Virasoro blocks with non-negative integer $q$-expansion coefficients, additionally the $S$-transformation matrix $M(S)$ that acts on the blocks should satisfy some integrality conditions coming from Verlinde's formula \cite{Verlinde1988}. Then we get candidates for MLDEs for characters and will be able to study sphere-torus correspondences more generally(if they exist) than the proposed one in certain types of theories.

\noindent\underline{\it 4. Infinite number of Virasoro blocks}: In unitary CFTs with $c>1$ whose spectra necessarily contain infinite number of Virasoro primaries, the conformal block expansions of some (crossing symmetric)correlators are likely to have an infinite number of Virasoro blocks as well. Construction of MLDEs for such blocks are out of the scope of the present program. However, such blocks may be expressed in terms of eigen functions of Laplace-Beltrami operator on the upper half plane with some choice of metric, generalising the ideas in \cite{BCFMP2021, KMP2021, BC2022, HMN2022}.

It will be also interesting to study MLDEs in the other possible cases given in Tab. \ref{tab:stabilizer}. For this, one need to generalise the code in \cite{CHKL2017} for analytical and/or numerical computations of the $q$-expansion coefficients in the $H$-functions adapted in those cases. When all the operators are distinct, we take a preliminary attempt as discussed at the end of appendix \ref{app:Hfunction}. It will be also interesting to consider higher order MLDEs and to generalize our construction for other type of conformal blocks. Finally, we should mention that the discrete values in \eqref{eq:discretealpha3} require an explanation in the alternative viewpoint on $q$-expansions of Virasoro blocks provided in \cite{MSZ2015} on top of the non-negativity of the $q$-expansion coefficients of pillow blocks.

\section*{Acknowledgements}

We would like to thank Sujay K. Ashok and Dileep P. Jatkar for useful comments on the draft.

\appendix
\numberwithin{equation}{section}

\section{Zamolodchikov's $H$-function}
\label{app:Hfunction}

The Zamolodchikov's $H$-functions are related to the Virasoro blocks, see  \eqref{eq:VBlockH} . The $q$-expansion coefficients in the $H$-functions can be obtained using Zamolodchikov's elliptic recursion formula \cite{Zamolodchikov1987}. We discuss relevant properties of these coefficients in this appendix. We also obtain the $q$-expansion coefficients of the Virasoro blocks.

To set up the recursion relation, we suitably adapt the parameterizations from \cite{CHKL2017} for our correlator $\langle O_1(\infty) O_2(1) O_3(x) O_4(0) \rangle$ as follows.\footnote{In contrast to us, \cite{CHKL2017} considers $\langle O_1(0) O_2(x) O_3(1) O_4(\infty) \rangle$. And, as discussed in appendix \ref{app:qexp}, $x$ is the inverse elliptic nome.}
\be
\begin{aligned}
  &c=1+6\big(b+\frac{1}{b}\big)^2,\qquad h_a=\frac{1}{4}\big(b+\frac{1}{b}\big)^2-\lambda_a^2\,, \\
  &h_{mn}=\frac{1}{4}\big(b+\frac{1}{b}\big)^2-\lambda_{mn}^2,\qquad \lambda_{mn}=\frac{1}{2}\big(\frac{m}{b}+nb\big)\,, \\
  &R_{mn}=2\prod_{p,q}(\lambda_3+\lambda_4-\lambda_{pq})(\lambda_4-\lambda_3-\lambda_{pq})(\lambda_1+\lambda_2-\lambda_{pq})(\lambda_2-\lambda_1-\lambda_{pq}) {\prod_{r,s}}'\lambda_{rs}^{-1}\,,
\label{eq:chbparameter}
\end{aligned}
\ee
where $p,q,r,s$ run over
\be
\begin{aligned}
  &p=-m+1,-m+3,\dots,m-3,m-1,\quad q=-n+1,-n+3,\dots,n-3,n-1\,, \\
  &r=-m+1,-m+2,\dots,m,\quad s=-m+1,-m+2,\dots,m\,, \\
\end{aligned}
\ee
and the primed product means that $(r,s)=(0,0)$ and $(m,n)$ are excluded. The recursion relation for $H$ is given by
\be
  H(c,h_a,h_p,q)=1+\sum_{m,n=1,2,\dots}\frac{q^{mn}R_{mn}}{h_p-h_{mn}}H(c,h_a,h_{mn}+mn,q)\,.
\label{eq:EllipRecur}
\ee

Note that $R_{mn}$ is symmetric when we interchange both $\lambda_1\leftrightarrow\lambda_4$ and $\lambda_2\leftrightarrow\lambda_3$, as the following holds.
\be
  \prod_{p,q}(\lambda_4-\lambda_3-\lambda_{pq})(\lambda_2-\lambda_1-\lambda_{pq})=\prod_{p,q}(\lambda_1-\lambda_2-\lambda_{pq})(\lambda_3-\lambda_4-\lambda_{pq})\,.
\ee
Hence, \eqref{eq:EllipRecur} asserts that $H$ is invariant under the interchange: $h_1\leftrightarrow h_4,h_2\leftrightarrow h_3$.

One can obtain the higher coefficients $b_n$ from the knowledge of the lower ones by evaluating \eqref{eq:EllipRecur} iteratively. All the odd coefficients $b_n$ contain a factor of $R_{11}$. Therefore, they vanish when $h_1=h_2$ or $h_3=h_4$ \cite{CHKL2017,Perlmutter2015}.

In the case $h_1=h_2\equiv h_L,\ h_3=h_4\equiv h_R$, \cite{CHKL2017} implements Zamolochikov's recursion relation for $H$ in a companion Mathematica notebook. The code can be used for both numerical and analytical computations of the $q$-expansion coefficients $b_n$ in $H$ yielding the analytic even coefficients $b_{2n}$ in terms of $b,h_L,h_R,h_p$ with odd ones vanishing. In the present paper, we need the first few analytic even coefficients $f_{2n}$ in \eqref{eq:qexpVblockgen}. $f_{2n}$ are linearly related to $b_{2n}$ \eqref{eq:linrelfnbn}. Thus, we use the above code to get the first few analytic coefficients $b_{2n}$, and replace the parameter $b$ in terms of $c$. Then, coding the relation \eqref{eq:VBHeta}, we get the required even coefficients of the Virasoro blocks. Below we present the first few. The higher analytic coefficients are very lengthy. For the vacuum block we have
\begin{align}
  f_2&=b_2+\frac{1}{2}\left(1-c+16 h_L+16 h_R\right) = \frac{512 h_L h_R}{c}-8 \left(h_L+h_R\right)\,, \nn \\
  f_4&=b_4+\frac{1}{8} \left(1-c+16 h_L+16 h_R\right) \left(4 b_2+7-c+16 h_L+16 h_R\right) \nn \\
  &=\frac{4}{c (5 c+22)}\Big[8 h_L^2 \left(640 h_R \left(2-c+32 h_R\right)+c (5 c+22)\right)+c (5 c+22) h_R \left(8 h_R-1\right) \nn \\
  &\qquad\qquad+h_L \left(16 h_R \left(128-320 (c-2) h_R+c (5 c-42)\right)-c (5 c+22)\right)\Big]\,.
\label{eq:fcoeffsVVB}
\end{align}
%
\begin{align}
f_6=&\frac{1}{3 c (-1+2 c) (22+5 c) (68+7 c)}\times \nn \\
&{\left(32 \left(-c (-1+2 c) (22+5 c) (68+7 c) h_L \left(1+h_L \left(-3+8 h_L\right)\right)+\right.\right.} \nn \\
&{\left(-c (-1+2 c) (22+5 c) (68+7 c)+2 (-23296+c (155320+3 c (12322+c (5113+70 c)))) h_L-\right.} \nn \\
&{24 (68928+c (35640+c (46346+c (-279+70 c)))) h_L^2+} \nn \\
&{\left.512 (376+c (15494+3 c (57+70 c))) h_L^3\right) h_R-} \nn \\
&{3 \left(-c (-1+2 c) (22+5 c) (68+7 c)+8 (68928+c (35640+c (46346+c (-279+70 c)))) h_L-\right.} \nn \\
&{\left.1024 (9256+c (9538+c (-839+70 c))) h_L^2+32768 (876+c (-251+70 c)) h_L^3\right) h_R^2-} \nn \\
&{8 \left(c (-1+2 c) (22+5 c) (68+7 c)-64 (376+c (15494+3 c (57+70 c))) h_L+\right.} \nn \\
&{\left.\left.\left.12288 (876+c (-251+70 c)) h_L^2-262144 (29+70 c) h_L^3\right) h_R^3\right)\right)}\,.
\label{eq:f6coeffVVB}
\end{align}
%
{\small
\begin{align}
f_8=&\frac{1}{3 c (-1+2 c) (46+3 c) (3+5 c) (22+5 c) (68+7 c)}\times \nn \\
&{\left(2 \left(c (-1+2 c) (46+3 c) (3+5 c) (22+5 c) (68+7 c) h_L \left(-3+4 h_L \left(35+16 h_L \left(-3+4 h_L\right)\right)\right)+\right.\right.} \nn \\
&{(-3 c (-1+2 c) (46+3 c) (3+5 c) (22+5 c) (68+7 c)+} \nn \\
&{8 (-15630336+c (93118128+c (2402116+c (-65081200+c (3581843+5 c (-163753+7350 c)))))) h_L-} \nn \\
&{64 (82286592+c (-121751536+c (-129927860+c (56360240+c (-4282919+15 c (72119+630 c)))))) } \nn \\
&{h_L^2+1024 (-7839360+c (-97857360+c (-2557588+c (-10723064+5 c (427349+c (3181+210 c)))))) } \nn \\
&{\left.h_L^3-65536 (-42864+c (-2709332+c (-1710448+c (259441+5 c (2173+210 c))))) h_L^4\right) h_R+} \nn \\
&{4 (35 c (-1+2 c) (46+3 c) (3+5 c) (22+5 c) (68+7 c)-} \nn \\
&{16 (82286592+c (-121751536+c (-129927860+c (56360240+c (-4282919+15 c (72119+630 c)))))) } \nn \\
&{h_L+} \nn \\
&{128 (-213555712+c (66109744+3 c (76581868+c (-7652528+5 c (1019645+c (2173+210 c)))))) h_L^2-} \nn \\
&{16384 (-13406640+c (-5758900+c (-1753868+15 c (195457+c (-2363+210 c))))) h_L^3+} \nn \\
&{\left.524288 (-831704+c (-410542+3 c (180603+25 c (-19+42 c)))) h_L^4\right) h_R^2+} \nn \\
&{64 (-3 c (-1+2 c) (46+3 c) (3+5 c) (22+5 c) (68+7 c)+} \nn \\
&{16 (-7839360+c (-97857360+c (-2557588+c (-10723064+5 c (427349+c (3181+210 c)))))) h_L-} \nn \\
&{1024 (-13406640+c (-5758900+c (-1753868+15 c (195457+c (-2363+210 c))))) h_L^2+} \nn \\
&{196608 (-386760+c (368134+5 c (50751+c (-2783+210 c)))) h_L^3-} \nn \\
&{\left.4194304 (-10686+c (24559+5 c (-599+210 c))) h_L^4\right) h_R^3+} \nn \\
&{256 (c (-1+2 c) (46+3 c) (3+5 c) (22+5 c) (68+7 c)-} \nn \\
&{256 (-42864+c (-2709332+c (-1710448+c (259441+5 c (2173+210 c))))) h_L+} \nn \\
&{8192 (-831704+c (-410542+3 c (180603+25 c (-19+42 c)))) h_L^2-} \nn \\
&{\left.\left.\left.1048576 (-10686+c (24559+5 c (-599+210 c))) h_L^3+16777216 (-251+5 c (661+210 c)) h_L^4\right) h_R^4\right)\right)}\,.
\label{eq:f8coeffVVB}
\end{align}
}
And, for the excited block corresponding to dimension $h_p$ we have
\be
\small
  f_2=\frac{4 h_p \left(3 c-2 (c-1) h_p\right)-8 h_L \left(2 h_p \left(c-8 h_p-64 h_R+11\right)+c-64 h_R\right)-8 h_R \left(2 h_p \left(c-8 h_p+11\right)+c\right)}{2 h_p \left(c+8 h_p-5\right)+c}\,.
\label{eq:fcoeffsVEB}
\ee
{\small
\begin{align}
  f_4=&\frac{2}{\left(c+8 h_p-1\right) \left(5 c \left(2 h_p+3\right)+2 \left(h_p-1\right) \left(8 h_p-33\right)\right) \left(2 c h_p+c+2 h_p \left(8 h_p-5\right)\right)}\times \nn \\
  &{\left(4 \left(-1+h_p\right) h_p \left(7 h_p \left(-13+4 h_p\right) \left(-3+8 h_p\right)-1206 h_R+64 h_p \left(273-386 h_p+64 h_p^2\right)
h_R+\right.\right.} \nn \\
&{\left.16 \left(-645+8 h_p \left(487+8 h_p \left(-43+8 h_p\right)\right)\right) h_R^2\right)+} \nn \\
&{64 h_L^2 \left(\left(-1+h_p\right) h_p \left(-645+8 h_p \left(487+8 h_p \left(-43+8 h_p\right)\right)\right)+\right.} \nn \\
&{\left.64 \left(-15+h_p \left(161+2 h_p \left(191+64 \left(-7+h_p\right) h_p\right)\right)\right) h_R+1024 \left(-15+4 h_p \left(26+h_p \left(-19+8
h_p\right)\right)\right) h_R^2\right)+} \nn \\
&{8 h_L \left(\left(-1+h_p\right) h_p \left(-603+32 h_p \left(273-386 h_p+64 h_p^2\right)\right)+\right.} \nn \\
&{16 \left(-96+h_p \left(1013+h_p \left(-6829+8 h_p \left(1307+8 h_p \left(-39+8 h_p\right)\right)\right)\right)\right) h_R+} \nn \\
&{\left.512 \left(-15+h_p \left(161+2 h_p \left(191+64 \left(-7+h_p\right) h_p\right)\right)\right) h_R^2\right)+} \nn \\
&{5 c^3 \left(16 h_L^2 \left(3+4 h_p \left(2+h_p\right)\right)+h_p \left(1+2 h_p\right) \left(9+2 h_p \left(-7+4 h_p\right)\right)-6 h_R+8 h_p
\left(-20+h_p \left(-1+8 h_p\right)\right) h_R+\right.} \nn \\
&{\left.16 \left(1+2 h_p\right) \left(3+2 h_p\right) h_R^2+2 h_L \left(3+2 h_p\right) \left(-1+16 h_R+2 h_p \left(-13+8 h_p+16 h_R\right)\right)\right)-} \nn \\
&{c^2 \left(h_p \left(-153-4 h_p \left(-484+h_p \left(919+4 h_p \left(-101+8 h_p\right)\right)\right)\right)+\right.} \nn \\
&{2 \left(51+4 h_p \left(-83+h_p \left(33+16 h_p \left(25+16 h_p\right)\right)\right)\right) h_R+16 \left(-51+4 h_p \left(-235+h_p \left(7+72
h_p\right)\right)\right) h_R^2+} \nn \\
&{2 h_L \left(51+4 h_p \left(-83+h_p \left(33+16 h_p \left(25+16 h_p\right)\right)\right)+2256 h_R+64 h_p \left(-467+h_p \left(71+232 h_p\right)\right)
h_R+\right.} \nn \\
&{\left.\left.5120 \left(3+4 h_p \left(2+h_p\right)\right) h_R^2\right)+16 h_L^2 \left(-51+1920 h_R+4 h_p \left(-235+7 h_p+72 h_p^2+640 \left(2+h_p\right)
h_R\right)\right)\right)+} \nn \\
&{2 c \left(h_p \left(-99+2 h_p \left(1238+h_p \left(-2771+4 \left(499-64 h_p\right) h_p\right)\right)\right)+66 h_R-\right.} \nn \\
&{4 h_p \left(748+h_p \left(-6929+8 h_p \left(1415+8 h_p \left(-73+8 h_p\right)\right)\right)\right) h_R+} \nn \\
&{16 \left(-33+2 h_p \left(-728+h_p \left(-353+16 h_p \left(73+12 h_p\right)\right)\right)\right) h_R^2+} \nn \\
&{2 h_L \left(33-2 h_p \left(748+h_p \left(-6929+8 h_p \left(1415+8 h_p \left(-73+8 h_p\right)\right)\right)\right)+4080 h_R-\right.} \nn \\
&{\left.32 h_p \left(176+h_p \left(369+16 h_p \left(-189+4 h_p\right)\right)\right) h_R+512 \left(45+2 h_p \left(-177+68 h_p+64 h_p^2\right)\right)
h_R^2\right)+} \nn \\
&{16 h_L^2 \left(-33+2880 h_R+2 \left(h_p \left(-728+h_p \left(-353+16 h_p \left(73+12 h_p\right)\right)\right)+\right.\right.} \nn \\
&{\left.\left.64 h_p \left(-177+68 h_p+64 h_p^2\right) h_R+5120 \left(3+4 h_p \left(2+h_p\right)\right) h_R^2\right)\right)}\,.
\end{align}
}
It is straightforward to check the symmetry of the above coefficients under $h_L\leftrightarrow h_R$, as previously argued.

The aforesaid code in \cite{CHKL2017} cannot be used for the calculations of the coefficients $b_n$ when all the external operators are distinct, and needs to be generalised. In a preliminary analysis, coding \eqref{eq:EllipRecur} for the general case, we have calculated the first few coefficients $b_n$, which upon setting $h_1=h_2,\ h_3=h_4$ reproduce above results. Here, we present the first few analytic coefficients $b_n$ in the general case. The higher analytic coefficients are very lengthy.
\be
  b_1=\frac{8 (h_1-h_2) (h_3-h_4)}{h_p}\,.
\ee
%
{\small
\begin{align}
b_2=&\frac{1}{2 h_p \left(c+2 h_p \left(-5+c+8 h_p\right)\right)}\times \nn \\
&{\left(128 c \left(h_1-h_2\right){}^2 \left(h_3-h_4\right){}^2+\right.} \nn \\
&{((-1+c) c+} \nn \\
&{16 \left(-h_2 \left(c+48 h_3^2+16 h_4 \left(-1+3 h_4\right)-16 h_3 \left(1+6 h_4\right)\right)+c \left(4 h_3^2+h_4 \left(-1+4 h_4\right)-h_3
\left(1+8 h_4\right)\right)+\right.} \nn \\
&{4 h_1^2 \left(c+16 h_3^2+4 h_4 \left(-3+4 h_4\right)-4 h_3 \left(3+8 h_4\right)\right)+4 h_2^2 \left(c+16 h_3^2+4 h_4 \left(-3+4 h_4\right)-4
h_3 \left(3+8 h_4\right)\right)+} \nn \\
&{\left.\left.h_1 \left(-c+16 h_3-48 \left(h_3-h_4\right){}^2+16 h_4-8 h_2 \left(c+16 h_3^2+4 h_4 \left(-3+4 h_4\right)-4 h_3 \left(3+8 h_4\right)\right)\right)\right)\right)
h_p+} \nn \\
&{2 \left(5+6 c+c^2+64 h_1^2+64 h_2^2-48 h_3-16 c h_3+64 h_3^2-16 h_2 \left(3+c-16 h_3-16 h_4\right)-\right.} \nn \\
&{\left.\left.16 h_1 \left(3+c+8 h_2-16 h_3-16 h_4\right)-16 \left(3+c+8 h_3\right) h_4+64 h_4^2\right) h_p^2\right)}\,.
\end{align}
}
%
{\small
\begin{align}
b_3=&\frac{4 \left(h_1-h_2\right) \left(h_3-h_4\right)}{3 h_p \left(2+c+h_p \left(-7+c+3 h_p\right)\right) \left(c+2 h_p \left(-5+c+8 h_p\right)\right)}\times \nn \\
&{\left(c (2+c) \left(5+3 c+64 h_3^2+16 h_4 \left(-3+4 h_4\right)-16 h_3 \left(3+8 h_4\right)\right)+\right.} \nn \\
&{\left(-100+c \left(19+48 c+9 c^2\right)+64 (-20+c (-1+3 c)) h_3^2+16 h_4 \left(60-9 c (1+c)+4 (-20+c (-1+3 c)) h_4\right)-\right.} \nn \\
&{\left.16 h_3 \left(-60+9 c (1+c)+8 (-20+c (-1+3 c)) h_4\right)\right) h_p+} \nn \\
&{3 \left(-310+c (-5+c (49+2 c))+64 (6+5 c) h_3^2+16 h_4 \left(42-c (27+2 c)+4 (6+5 c) h_4\right)-\right.} \nn \\
&{\left.16 h_3 \left(-42+c (27+2 c)+8 (6+5 c) h_4\right)\right) h_p^2+} \nn \\
&{6 \left(167+c (154+3 c)+192 h_3^2-48 \left(19+c-4 h_4\right) h_4-48 h_3 \left(19+c+8 h_4\right)\right) h_p^3+1152 h_p^4+} \nn \\
&{48 h_2 \left(-c (2+c)-16 h_3^2 \left(c+h_p \left(-7+3 c+9 h_p\right)\right)-16 h_4^2 \left(c+h_p \left(-7+3 c+9 h_p\right)\right)+\right.} \nn \\
&{h_p \left(20-3 c (1+c)-h_p \left(-42+c (27+2 c)+6 (19+c) h_p\right)\right)+} \nn \\
&{16 h_4 \left(c+h_p \left(3 (-3+c)+h_p \left(9+2 c+6 h_p\right)\right)\right)+} \nn \\
&{\left.16 h_3 \left(c-9 h_p+h_p \left(3+2 h_p\right) \left(c+3 h_p\right)+2 h_4 \left(c+h_p \left(-7+3 c+9 h_p\right)\right)\right)\right)+} \nn \\
&{64 h_1^2 \left(c (2+c)+h_p \left(-20+c (-1+3 c)+3 h_p \left(6+5 c+6 h_p\right)\right)-12 h_4 \left(c+h_p \left(-7+3 c+9 h_p\right)\right)+\right.} \nn \\
&{2 h_3^2 \left(c (8+c)+h_p \left(-26+11 c+24 h_p\right)\right)+2 h_4^2 \left(c (8+c)+h_p \left(-26+11 c+24 h_p\right)\right)+} \nn \\
&{\left.4 h_3 \left(-3 \left(c+h_p \left(-7+3 c+9 h_p\right)\right)-h_4 \left(c (8+c)+h_p \left(-26+11 c+24 h_p\right)\right)\right)\right)+} \nn \\
&{64 h_2^2 \left(c (2+c)+h_p \left(-20+c (-1+3 c)+3 h_p \left(6+5 c+6 h_p\right)\right)-12 h_4 \left(c+h_p \left(-7+3 c+9 h_p\right)\right)+\right.} \nn \\
&{2 h_3^2 \left(c (8+c)+h_p \left(-26+11 c+24 h_p\right)\right)+2 h_4^2 \left(c (8+c)+h_p \left(-26+11 c+24 h_p\right)\right)+} \nn \\
&{\left.4 h_3 \left(-3 \left(c+h_p \left(-7+3 c+9 h_p\right)\right)-h_4 \left(c (8+c)+h_p \left(-26+11 c+24 h_p\right)\right)\right)\right)+} \nn \\
&{16 h_1 \left(-3 c \left(2+c+16 \left(-1+h_4\right) h_4\right)-3 \left(-20+3 c (1+c)+16 h_4 \left(9-3 c+(-7+3 c) h_4\right)\right) h_p+\right.} \nn \\
&{3 \left(42-c (27+2 c)+16 \left(9+2 c-9 h_4\right) h_4\right) h_p^2-18 \left(19+c-16 h_4\right) h_p^3-48 h_3^2 \left(c+h_p \left(-7+3 c+9 h_p\right)\right)+} \nn \\
&{48 h_3 \left(c-9 h_p+h_p \left(3+2 h_p\right) \left(c+3 h_p\right)+2 h_4 \left(c+h_p \left(-7+3 c+9 h_p\right)\right)\right)+} \nn \\
&{8 h_2 \left(-c (2+c)+h_p \left(20+c-3 c^2-3 h_p \left(6+5 c+6 h_p\right)\right)+12 h_4 \left(c+h_p \left(-7+3 c+9 h_p\right)\right)-\right.} \nn \\
&{2 h_3^2 \left(c (8+c)+h_p \left(-26+11 c+24 h_p\right)\right)-2 h_4^2 \left(c (8+c)+h_p \left(-26+11 c+24 h_p\right)\right)+} \nn \\
&{\left.\left.\left.4 h_3 \left(3 \left(c+h_p \left(-7+3 c+9 h_p\right)\right)+h_4 \left(c (8+c)+h_p \left(-26+11 c+24 h_p\right)\right)\right)\right)\right)\right)}\,.
\end{align}
}

\section{$q$-expansion of certain functions}
\label{app:qexp}

The $q$-expansions of the functions $\lambda(\tau),\theta_2(\tau),\theta_3(\tau),\theta_4(\tau),\eta(\tau),E_2(\tau),E_4(\tau),E_6(\tau)$ are used in this paper. Exact formulae for the $q$-expansions of $\theta_r$ and $\eta$ are given by
\begin{align}
  &\theta_2(\tau)=2\sum_{n=1,3,5,\dots} q^{n^2/4}\,, \nn \\
  &\theta_3(\tau)=1+2\sum_{n=1}^{\infty} q^{n^2}\,, \nn \\
  &\theta_4(\tau)=1+2\sum_{n=1}^{\infty} (-1)^n q^{n^2}\,, \nn \\
  &\eta(\tau)=\sum_{n=0,\mp1,\mp2,\dots}(-1)^n q^{(6n+1)^2/12}\,.
\label{eq:qexpThetasEta}
\end{align}
For the functions $\lambda(\tau),E_2(\tau),E_4(\tau),E_6(\tau)$, below we present formulae in terms of some Mathematica library functions. Then, it is straightforward to write Mathematica routines to obtain their $q$-expansion coefficients to any desired order $q^n$.

First note that the local inverse of \eqref{eq:mapUHPx} is given by\cite{Zamolodchikov1987,Handbooks}
\begin{align}
  &\tau(\lambda)=i\frac{K(1-\lambda)}{K(\lambda)}\,, \nn \\
  &K(\lambda)\equiv\int_0^1\frac{dt}{\sqrt{\left(1-t^2\right) \left(1-\lambda  t^2\right)}}=\frac{1}{2} \pi  \, _2F_1\left(\frac{1}{2},\frac{1}{2};1;\lambda \right)\,,
\end{align}
where $K$ denotes the complete elliptic integral of the first kind. The elliptic nome is given by
\be
  q(\lambda)=e^{i\pi\tau(\lambda)}\,.
\ee
The elliptic nome and its inverse are Mathematica library functions, EllipticNomeQ[$\lambda$] and InverseEllipticNomeQ[$q$], respectively, with the latter providing the elliptic lambda function $\lambda$ as a function of $q$.\footnote{On the other hand, the Mathematica library function ModularLambda[$\tau$] provides $\lambda(\tau)$.}

The library function EllipticTheta[$r, q$] provides the Jacobi theta functions $\theta_r,\ r=2,3,4$ as a function of $q$. As presented below, the Eisenstein series $E_2$ can be expressed in terms of the complete elliptic integral of the second kind $E$, the elliptic lambda function $\lambda$ and the Jacobi theta functions $\theta_r$, and $E_4,E_6$ in terms of $\theta_r$ \cite{Handbooks}. Using Mathematica library function EllipticE[$\lambda$] for $E(\lambda)$, EllipticE[InverseEllipticNomeQ[$q$]] provides its $q$-dependency.
\begin{align}
  &E_2=\frac{6}{\pi}E(\lambda)\theta_3^2-\theta_3^4-\theta_4^4,\quad E(\lambda)\equiv\int_0^1\sqrt{\frac{1-\lambda  t^2}{1-t^2}} dt\,, \nn \\
  &E_4=\frac{1}{2}(\theta_2^8+\theta_3^8+\theta_4^8)\,, \nn \\
  &E_6=\frac{1}{2} \left(\theta _3^{12}+\theta _4^{12}-3 \theta _2^8 (\theta _3^4+\theta _4^4\right))\,.
\end{align}

Now using Mathematica, the coefficients of the $q$-expansions of $\lambda(\tau),E_2(\tau),E_4(\tau),E_6(\tau)$ can be obtained to any desired order $q^n$. Here, we present the first few terms:
\begin{align}
  &\lambda(\tau)=16 q - 128 q^2 + 704 q^3 - 3072 q^4 + 11488 q^5 - 38400 q^6 + 
 117632 q^7 - 335872 q^8+\cdots\,, \nn \\
  &E_2(\tau)=1 - 24 q^2 - 72 q^4 - 96 q^6 - 168 q^8+\cdots\,, \nn \\
  &E_4(\tau)=1 + 240 q^2 + 2160 q^4 + 6720 q^6 + 17520 q^8+\cdots\,, \nn \\
  &E_6(\tau)=1 - 504 q^2 - 16632 q^4 - 122976 q^6 - 532728 q^8+\cdots\,.
\end{align}

\section{Closed form solutions}
\label{app:tautox}

In this appendix, we consider the second order MLDE w.r.t. $\Gamma(2)$ in \eqref{eq:2ndmlde} and provide 2 solutions in closed form. The solutions are so chosen that they shall be directly identified -- whenever possible on the parameter space $\vec{\alpha}$ -- with conformal blocks from their $q\to0$ behaviour.

Under the change of variable $x=\lambda(\tau)$, \eqref{eq:2ndmlde} leads to\footnote{In order to obtain closed form solutions to MLDEs of orders 2 and 3 for the characters, \cite{MMS1988b} made a variable change from the torus parameter $\tau$ to an auxiliary space $x=\lambda(\tau)$. Here, we follow the same path, except in our case $x$-space should be interpreted as the cross-ratio space.}
\be
  x^2(1-x)^2\frac{d^2\vec{F}}{dx^2}+x(1-x)\big(1-2 \alpha _2-(2 \alpha _1+1) x\big)\frac{d\vec{F}}{dx}-(\alpha_3x^2+\alpha_4x+\alpha_5)\vec{F}=0\,,
\label{eq:2ndodex}
\ee
where we have used the following relations.
\be
  \frac{d\lambda}{d\tau}=i\pi\lambda(1-\lambda)\theta_3^4,\quad E_2=-(2\lambda-1)\theta_3^4+\frac{3}{i\pi\theta_3^4}\frac{d}{d\tau}\theta_3^4\,.
\ee
\eqref{eq:2ndodex} is the most general second order linear differential equation with 3 regular singular points viz. at $x=0,1,\infty$.\footnote{In order for $\infty$ to be a regular singular point of a differential equation given below, $\lim_{y\to0}yP(\frac{1}{y})$ and $\lim_{y\to0}y^2Q(\frac{1}{y})$ should be finite. Thus, $P$ should be a linear while $Q$ a quadratic polynomial in $x$. \eqref{eq:2ndodex} has the most general such $P,Q$.
$$\frac{d^2f}{dx^2}+\frac{P(x)}{x(x-1)}\frac{df}{dx}+\frac{Q(x)}{x^2(x-1)^2}f=0\ .$$} Therefore, the solutions can be obtained in terms of Gauss hypergeometric function as
\begin{align}
  &F_{p(1)}(x)=x^{\mu_{-}}(1-x)^{\nu_{-}}\, _2F_1(a_{-},b_{-};c_{-};x),\quad F_{p(2)}(x)=x^{\mu_{+}}(1-x)^{\nu_{+}}\, _2F_1(a_{+},b_{+};c_{+};x)\,, \nn \\
  &\mu_{\mp}=\alpha _2 \mp \sqrt{\alpha _2^2+\alpha _5}\,, \nn \\
  &\nu_{\mp}=\frac{1}{2}- \alpha _1- \alpha _2 \mp \sqrt{\big(\alpha _1+\alpha _2-\frac{1}{2}\big)^2+ (\alpha _3+\alpha _4+\alpha _5)}\,, \nn \\
  &a_{\mp}=\frac{1}{2} \pm \sqrt{\alpha _1^2+\alpha _3} \mp \sqrt{\alpha _2^2+\alpha _5} \mp \sqrt{\big(\alpha _1+\alpha _2-\frac{1}{2}\big)^2+ (\alpha _3+\alpha _4+\alpha _5)}\,, \nn \\
  &b_{\mp}=\frac{1}{2} \mp \sqrt{\alpha _1^2+\alpha _3} \mp \sqrt{\alpha _2^2+\alpha _5} \mp \sqrt{\big(\alpha _1+\alpha _2-\frac{1}{2}\big)^2+ (\alpha _3+\alpha _4+\alpha _5)}\,, \nn \\
  &c_{\mp}=1 \mp 2 \sqrt{\alpha _2^2+\alpha _5}\,.
\label{eq:closedsolshol}
\end{align}
It is straightforward to obtain the actions of $T,S$ on the above solutions using the following identities \cite{Handbooks}.
\begin{align}
  &_2F_1(a,b;c;x)=(1-x)^{c-a-b}\, _2F_1(c-a,c-b;c;x)\,, \nn \\
  &_2F_1(a,b;c;T\cdot x)=(1-x)^{a}\, _2F_1(a,c-b;c;x)=(1-x)^{b}\, _2F_1(c-a,b;c;x)\,, \nn \\
  &_2F_1(a,b;c;S\cdot x)=\frac{\Gamma(c)\Gamma(c-a-b)}{\Gamma(c-a)\Gamma(c-b)}\, _2F_1(a,b;a+b-c+1;x) \nn \\
  &\qquad\qquad\quad+\frac{\Gamma(c)\Gamma(a+b-c)}{\Gamma(a)\Gamma(b)}x^{c-a-b}\, _2F_1(c-a,c-b;c-a-b+1;x)\,, \nn \\
  &_2F_1(a,b;c;S\cdot x)=\frac{\Gamma(c)\Gamma(a+b-c)}{\Gamma(a)\Gamma(b)}x^{c-a-b} (1 - x)^{1-c}\, _2F_1(1-b,1-a; 1+ c-a-b; x) \nn \\
  &\qquad\qquad\quad+\frac{\Gamma(c)\Gamma(c-a-b)}{\Gamma(c-a)\Gamma(c-b)} (1-x)^{1-c}\, _2F_1(1+ b-c,1+a -c; 1+a+b-c;x)\,.
\label{eq:hypergeoid}
\end{align}
In particular, we compute that\footnote{We also use the relation $\Gamma(1–z)\Gamma(z)=\pi\csc\pi z$.}
\begin{align}
  &F(ST^2S\cdot x)=\frac{\left(e^{2 i \pi  (a+c)}+e^{2 i \pi  (b+c)}-e^{2 i \pi  (a+b)}-e^{2 i \pi  c}\right) e^{-2 i \pi  (a+b-\nu )}}{e^{2 i \pi  c}-1} F(x) \nn \\
  &-\frac{2 i \pi  \Gamma (c-1) \Gamma (c) e^{-i \pi  (a+b-c-2 \nu )}}{\Gamma (a) \Gamma (b) \Gamma (c-a) \Gamma (c-b)} x^{1-c+\mu} (1-x)^{c-a-b+\nu } \, _2F_1(1-a,1-b;2-c;x)\,, \nn \\[10pt]
  &F(x)=x^{\mu } (1-x)^{\nu } \, _2F_1(a,b;c;x)\,.
\label{eq:actionSTTSonF}
\end{align}
And, note that the values of $\{1-c+\mu,\ c-a-b+\nu,\ 1-a,\ 1-b,\ 2-c\}$ for any of the 2 solutions in \eqref{eq:closedsolshol} give the respective values of $\{\mu,\nu,a,b,c\}$ for the other solution.

\end{document}